\documentclass[12pt,preprint]{aastex}

\def\ergs{ergs s$^{\rm -1}$ }
\def\flux{ergs s$^{\rm -1}$ cm s$^{\rm -2}$ }

\def\msolar{\mbox{M$_{\odot}$} }

\def\ltsima{$\; \buildrel < \over \sim \;$}
\def\simlt{\lower.5ex\hbox{\ltsima}}            % < over ~
\def\gtsima{$\; \buildrel > \over \sim \;$}
\def\simgt{\lower.5ex\hbox{\gtsima}}            % > over ~

\def\beq{\begin{equation}}
\def\eeq{\end{equation}}
\def\beqary{\begin{eqnarray}}
\def\eeqary{\end{eqnarray}}
\newcommand{\eqref}[1]{(\ref{eq:#1})}

\begin{document}


%% \title{ Discovery of Diffuse Hard X-ray Emission from 
%%         the Massive Star-Forming Region NGC 6334}
 \title{ Investigation of Diffuse Hard X-ray Emission from 
         the Massive Star-Forming Region NGC 6334}


\author{Y. Ezoe\altaffilmark{1,2}, 
        M. Kokubun\altaffilmark{1}, 
        K. Makishima\altaffilmark{1,3},
        Y. Sekimoto\altaffilmark{4}, 
        K. Matsuzaki\altaffilmark{2}} 


\altaffiltext{1}{Department of Physics, University of Tokyo,
                 7-3-1 Hongo, Bunkyo-ku, Tokyo, Japan}
\altaffiltext{2}{The Institute of Space and Astronautical Science,
                 3-1-1 Yoshinodai, Sagamihara, Kanagawa 229-8510, Japan}
\altaffiltext{3}{RIKEN (The Institute of Physical and Chemical Research), 
                 2-1, Hirosawa, Wako, Saitama 351-0198, Japan}
\altaffiltext{4}{National Astronomical Observatory of Japan,
                 2-21-1 Osawa, Mitaka, Tokyo, 181-8588, Japan}


\begin{abstract}
{\it Chandra} ACIS--I data of the molecular cloud and HII region complex 
NGC 6334 were analyzed. The hard X-ray clumps detected with {\it ASCA} 
\citep{Sekimoto2000} were resolved into 792 point sources. After removing 
the point sources, an extended X-ray emission component was detected over 
a $5\times9$ pc$^2$ region, with the 0.5--8 keV absorption-corrected 
luminosity of 2$\times10^{33}$ \ergs. The contribution from faint point 
sources to this extended emission was estimated as at most $\sim20$\%, suggesting 
that most of the emission is diffuse in nature. The X-ray spectrum of the 
diffuse emission was observed to vary from place to place. 
In tenuous molecular cloud regions with hydrogen column density of 
$0.5\sim1\times10^{22}$ cm$^{-2}$, the spectrum can be represented 
by a thermal plasma model with temperatures of several keV. 
The spectrum in dense cloud cores exhibits harder continuum, 
together with higher absorption more than $\sim3\times10^{22}$ cm$^{-2}$. 
In some of such highly obscured regions, the spectrum show 
extremely hard continua equivalent to a photon index of $\sim1$, and 
favor non-thermal interpretation.
These results are discussed in the context of thermal and non-thermal
emissions, both powered by fast stellar winds from embedded young 
early-type stars through shock transitions.
\end{abstract}




\section{Introduction}
\label{sec:intro}

X-rays provide a powerful diagnostic tool of massive star-forming regions 
(MSFRs). X-ray photons penetrate dense gas and dust cores more deeply than 
optical and even near infrared lights. Not only serving as a probe, the X-ray 
emission itself provides evidence that some unexpected energetic processes 
are operating in MSFRs. Therefore, since the surprising discovery of X-ray 
emission from the Orion nebula \citep{Giacconi1974}, vigorous X-ray 
studies on MSFRs have been made with {\it Einstein}, {\it ROSAT}, and 
{\it ASCA}. Especially, the hard X-ray ($>$ 2 keV) imaging spectroscopy 
with {\it ASCA} has for the first time revealed several clumps in MFSRs, 
each showing a high temperature (3--10 keV) and a 
high luminosity (10$^{32-34}$ \ergs) 
(e.g., \citealt{Yamauchi1996,Hofner1997,Nakano2000,Sekimoto2000}).
Thanks to its superb angular resolution, these X-ray clumps in MSFRs have 
been resolved into hundreds of stars with {\it Chandra} 
(e.g., \citealt{Garmire2000, Feigelson2002a, 
Feigelson2003, Hofner2002, Kohno2002, Nakajima2003}). 

Besides individual stars at various evolutionary stages, {\it Chandra} first 
enabled a detailed examination of diffuse X-ray emission 
in MSFRs. Theoretically, such a phenomenon had long been expected 
both around individual OB stars \citep{Dyson1972,Castor1975,Weaver1977} as well as in
MSFRs \citep{Chevalier1985}. 
The reported detections of extended soft X-rays ($\lesssim$ 1 keV) 
associated with MSFRs before {\it Chandra}
(e.g., the Carina Nebula, \citealt{Seward1982}; 30 Doradus nebula in 
the Large Magelanic Cloud, \citealt{Wang1999})
were consistent with these predictions. 
However, even in Galactic MSFRs, it has remained difficult to date to 
quantify whether the extended X-ray emission really exists, because MSFRs are
usually too complex to be resolved by the past X-ray telescopes.

With {\it Chandra}, \citet{Townsley2003} indeed reported detections of 
diffuse soft X-ray emission in two nearby MSFRs, M17 
(at a distance of $D=$ 1.6 kpc from the Sun) 
and the Rosette Nebula ($D=$ 1.4 kpc). 
They concluded that at least the emission in M17 is diffuse, 
and explained the phenomenon
in terms of stellar-wind shock model \citep{Dyson1972,Castor1975, Weaver1977}. 
Its spectrum was very soft, represented by 
a two temperature plasma model ($kT=0.13+0.6$ keV).
Another {\it Chandra} result is the possible detection of 
diffuse hard X-ray emission from RCW 38 ($D=$ 1.7 kpc; 
\citealt{Wolk2002}), which could be of non-thermal origin
because a part of the emission show hard continua with a photon index 
of $1.3\sim1.6$.
Similar phenomena may have also been detected from more 
distant galactic MSFRs, including the Arches cluster 
\citep{Wang2002,Yusef-Zadeh2002}, the Quintuplet cluster 
\citep{Wang2002} near the Galactic central, and NGC 3603 at $D=$ 7 kpc 
\citep{Moffat2002}, although their larger distances make it difficult 
to remove, even with {\it Chandra}, contributions from unresolved point 
sources. 

Thus, the search for diffuse X-ray emission from MSFRs is becoming
a new topic explored with {\it Chandra}. In the present paper, we 
describe the analysis of {\it Chandra} data of NGC 6334, and report on 
the discovery 
of diffuse hard X-ray emission from this representative MSFR.

\section{Previous X-ray Results on NGC 6334}
\label{sec:ngc6334}

\begin{figure}[htbp]
%%{fig/ngc6334-allmap.tgif.eps}
\plotone{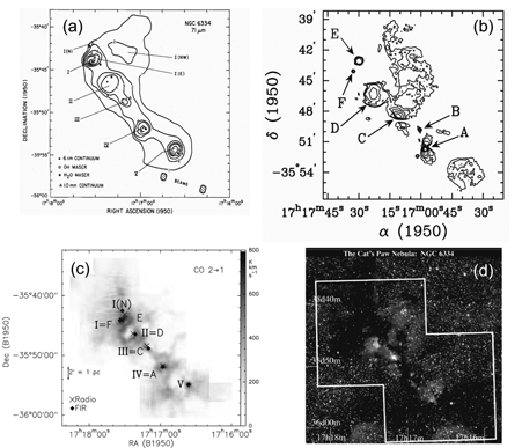}
\caption{Images of NGC 6334 in; (a) the FIR 71$\mu$m emission
        (\protect\citealt{Loughran1986});
         (b) radio 18 cm continuum emission (\protect\citealt{Sarma2000});
         (c) CO $J$=2-1 (\protect\citealt{Kraemer1999a});
         and (d) NIR (the 2$\mu$m All Sky Survey Atlas image).
         Roman numerals and Arabic characters designate the 
         FIR cores and HII regions, respectively.
         In the CO map, crosses and diamonds indicate
         the positions of the FIR cores and radio peaks.
         In the NIR map, three colors (blue, green
         and red) correspond to J (1.2$\mu$m), H (1.6$\mu$m), and
         K$_{\rm s}$ (2.2$\mu$m) bands, respectively, and squares 
         indicate the two fields-of-view of the {\it Chandra} observations.
         }
\label{fig:ngc6334-maps}
\end{figure}

NGC 6334 is a nearby ($D=$ 1.7 kpc yielding a plate scale of 
1$^{\prime\prime}=$0.01 pc; \citealt{Neckel1978}) MSFR,
with the bolometric luminosity reaching 
$L_{\rm bol} \sim 4\times 10^{39}$ ergs s$^{-1}$ \citep{Loughran1986} 
which is one of the highest of this class.
As shown in figure \ref{fig:ngc6334-maps}, it contains several 
star-forming sites defined in wide wavelength ranges.
Although designations of these sites depend on the observing
wavelength, we here follow two commonly used ones; 
the far-infrared (FIR) cores named I(N), I, II, III, IV, and V 
(figure \ref{fig:ngc6334-maps}a); and the radio sources named
A, C, D, E, and F (figure \ref{fig:ngc6334-maps}b). 
Not all of these are detected in both 
wavelength ranges (see \citealt{Kraemer1999a} table 4 for nomenclature).
Each site is known to be powered by one or more massive stars, 
either zero-age main sequence stars (ZAMS) or protostars 
(e.g., \citealt{Rodriguez1982,Harvey1983,Straw1989a}). 
These sites coincide with the dense molecular clouds 
(figure \ref{fig:ngc6334-maps}c),
the densest part of which appears as the central dark lane in the
near-infrared (NIR) map (figure \ref{fig:ngc6334-maps}d). 
The central FIR core (III) has no radio masers or outflows
that indicate early stages of stellar evolution, but hosts HII regions.
It is therefore thought to be most evolved. On the other hand, the FIR 
cores I, IV and V have outflows and masers, suggesting that the 
star formation has just started there. 
The core I (N) has no HII region but masers and an outflow, so that this 
core is regarded as one of the youngest massive star-forming sites.

In X-rays, NGC 6334 was first detected with the {\it Einstein} satellite
as 2E1717.1-3548 \citep{Harris1990}, which coincides in position with the FIR 
core III. With {\it ASCA}, \citet{Matsuzaki1999a}, \citet{Matsuzaki1999b}
and \citet{Sekimoto2000} 
found hard X-ray clumps associated with the FIR and radio sources; five FIR 
cores (I, II, III, IV, and V) were visible in the 2--10 keV band. The 
region-integrated spectrum exhibits a temperature of 9 keV, which is the
highest among the MSFRs observed with {\it ASCA}. The X-rays are absorbed by 
a large absorption column density ($N_{\rm H}\sim 1\times10^{22}$ cm$^{-2}$),
suggesting that the emission originates from
embedded young massive stars. The region-integrated 0.5--10 keV X-ray 
luminosity, 6$\times$10$^{33}$ \ergs, makes NGC 6334 one of the most X-ray
luminous MSFRs ever observed with {\it ASCA}. Thus, NGC 6334 is without doubt 
an attractive target to search for diffuse hard X-rays, and if detected, 
to investigate their properties and emission mechanism.

\section{Observation and Data Reduction}
\label{sec:observation}

We conducted two 40 ksec observations of NGC 6334 with {\it Chandra}, 
on 2002 August 31 and 2002 September 2. In order to cover the whole nebula, 
we placed the aimpoint at the J2000 co-ordinates 
of (17$^{\rm h}$ 20$^{\rm m}$ 53.\hspace*{-0.1cm}$^{\rm s}$45, 
$-$35$^{\rm d}$  47$^{\rm m}$ 19.\hspace*{-0.1cm}$^{\rm s}$33) 
and (17$^{\rm h}$ 20$^{\rm m}$ 00.\hspace*{-0.1cm}$^{\rm s}$44, 
$-$35$^{\rm d}$  56$^{\rm m}$ 22.\hspace*{-0.1cm}$^{\rm s}$26)
in the August and September observations, respectively. 
Hereafter we call the former ``north field'' and the latter ``south field'',
according to their declinations. The two fields of view partially overlap
(see figure \ref{fig:ngc6334-maps} d). The Advanced CCD 
Imaging Spectrometer (ACIS--I0, I1, I2, I3, S2, and S3) was utilized. 
In this paper, we analyze only the data from the four ACIS--I chips, 
because the point spread function (PSF) of the telescope 
becomes significantly broader 
at the positions of the ACIS--S chips.

We started data reduction with the Level 1 event files provided by the 
{\it Chandra} X-ray Center, and followed the standard data reduction procedures
using the CIAO (Chandra Interactive Analysis of Observations) software package 
version 2.3 and the calibration data base (caldb) version 2.18. We corrected 
pulse heights for the charge transfer inefficiency (CTI), and removed the 
ACIS pixel randomization to improve image quality. 
No background flares occurred in the north-field observation, while 
nearly half the south-field exposure was affected by them. 
We then conducted flare filtering, by excluding those portions where 
the 0.5--8 keV count rate was $>$1.2 times the quiescent average of the 
south-field observation 
($\sim$4$\times10^{-7}$ ct s$^{-1}$ pixel$^{-1}$). 
We have thus obtained 39.4 and 19.4 ksec exposure times for the north 
and south observations, respectively.

\begin{figure}[htbp]
%%{fig/csm_0.5-2_cmap.tgif.eps}
%%{fig/csm_2-8_cmap.tgif.eps}
\plottwo{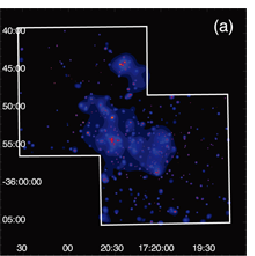}{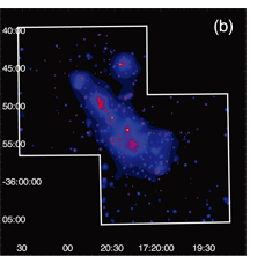}
\caption{Adaptively smoothed X-ray images of NGC 6334 in the
         (a) 0.5--2 and (b) 2--8 keV bands, displayed on the J2000
         coordinates. They are corrected for the exposure and vignetting,
         but the background is not subtracted.
         The CIAO program {\bf csmooth} is utilized.
         The two observed fields are merged together.
         The ACIS--I fields of view are shown in white lines.
         The intensity is plotted logarithmically from
         1.0$\times10^{-9}$ to 2.9$\times10^{-6}$
         counts s$^{-1}$ pixel$^{-1}$ cm$^{-2}$ in panel (a),
         while from
         1.8$\times10^{-9}$ to 2.9$\times10^{-6}$
         in panel (b).
        }
\label{fig:ngc6334-chandra-csmimage}
\end{figure}

Using these data sets, we created X-ray images in the 0.5--2 and 
2--8 keV bands, as shown in figure \ref{fig:ngc6334-chandra-csmimage}.
We observe a number of point sources and also indication of extended
structures. However, it is not clear at this stage whether they really 
represent diffuse emission or simply a sum of point sources which cannot 
be resolved even with {\it Chandra}.
Therefore, before analyzing the data for suggested extended emission,
we conduct point source analysis in the next section.

\section{Point Source Analysis}
\label{sec:ps-extract}

The {\bf wavdetect} program was utilized to detect point sources 
(\citealt{Freeman2002}). We used images in three bands (0.5--2, 2--8, 
and 0.5--8 keV) independently, in order not to miss very soft or 
strongly absorbed sources. The significance criterion and wavelet 
scales were set as 1$\times10^{-6}$ and 1 to 16 pixels in multiples
of $\sqrt{2}$, respectively. As a result, we have detected 449 and 
390 sources in the north and south fields, respectively. Among them,
42 sources in the north field have counterparts in the south field 
within the position-dependent angular resolution, approximately
defined in \cite{Feigelson2002a}.
Excluding these overlaps, the total source number becomes 792. 

We extracted events of the individual point sources using regions given 
by the {\bf wavdetect} program. The background was extracted around each 
source excluding other point sources, and then scaled to the on-source area.
The scaled background counts amount to $\sim$1-8 \% of the raw source counts
near the aimpoint, while $\sim$20-40 \% for sources at $8^{\prime}$.
After subtracting the background, the detected sources exhibit net signal
counts (hereafter netcounts) of $\sim$2 to 3000. Although faint sources 
less than $\simlt$5 netcounts have low significance ($\simlt2\sigma$), 
we here take them as possible point sources, in order not to include their 
counts into the potentially extended emission.
We then produced source-number histograms as a function of logarithm of 
netcounts, in 6 separate annular regions (0-2, 2-4, 4-6, 6-8, 8-10 
and $>$10 arcminites from the aimpoint) in each ACIS--I field. 
Reflecting a roughly power-law like source intensity distribution 
(see \S \ref{sec:ana:lumi} for details), the source number first 
increases toward lower counts, reaches a maximum, and then decreases
due to the sensitivity limit. We regard this maximum of the histogram
as representing the completeness limit of the point-source detection.
In both observations, the completeness limit estimated in this way 
monotonically increases with the angular distance from 
$\sim$10 to $\sim$30 netcounts, due to the PSF broadening.
When utilizing a typical count-to-flux conversion factor of 
$\sim2\times10^{-11}$ \flux (netcounts/s)$^{-1}$ obtained from spectral 
fitting of the 163 bright ($>30$ netcounts) sources, 10 netcounts 
corresponds to the 0.5--8 keV X-ray flux of $\sim0.5\times10^{-14}$ 
and $\sim1\times10^{-14}$ \flux, in the north and south fields, 
respectively. At an assumed distance of 1.7 kpc, these fluxes translate 
to $\sim2\times10^{30}$ and $\sim4\times10^{30}$ \ergs. 
Detailed point source analyses will be presented elsewhere 
(Ezoe et al. in preparation).

\begin{figure}[htbp]
%%{fig/mask_n.tgif.eps}
%%{fig/mask_s.tgif.eps}
\plottwo{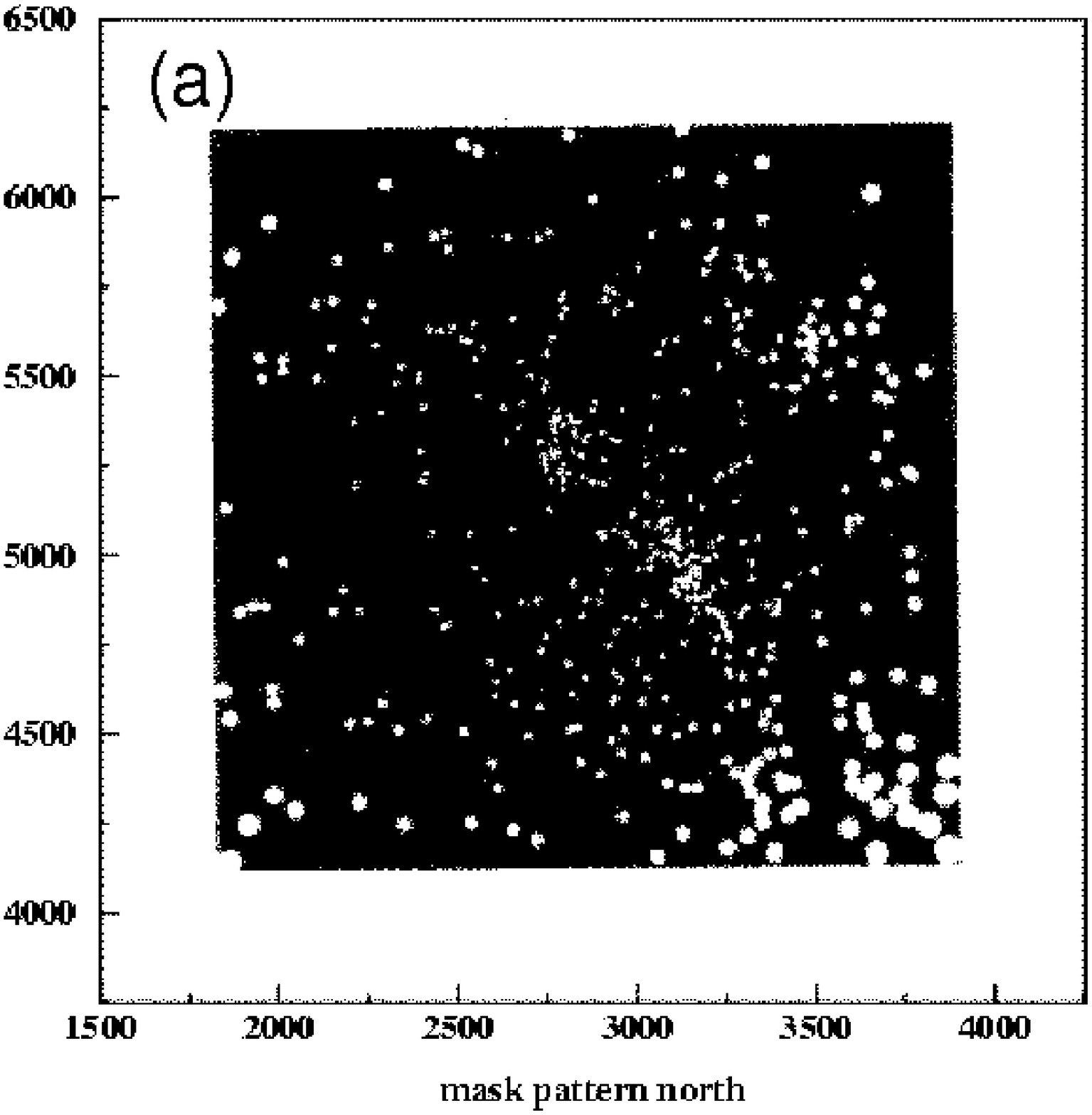}{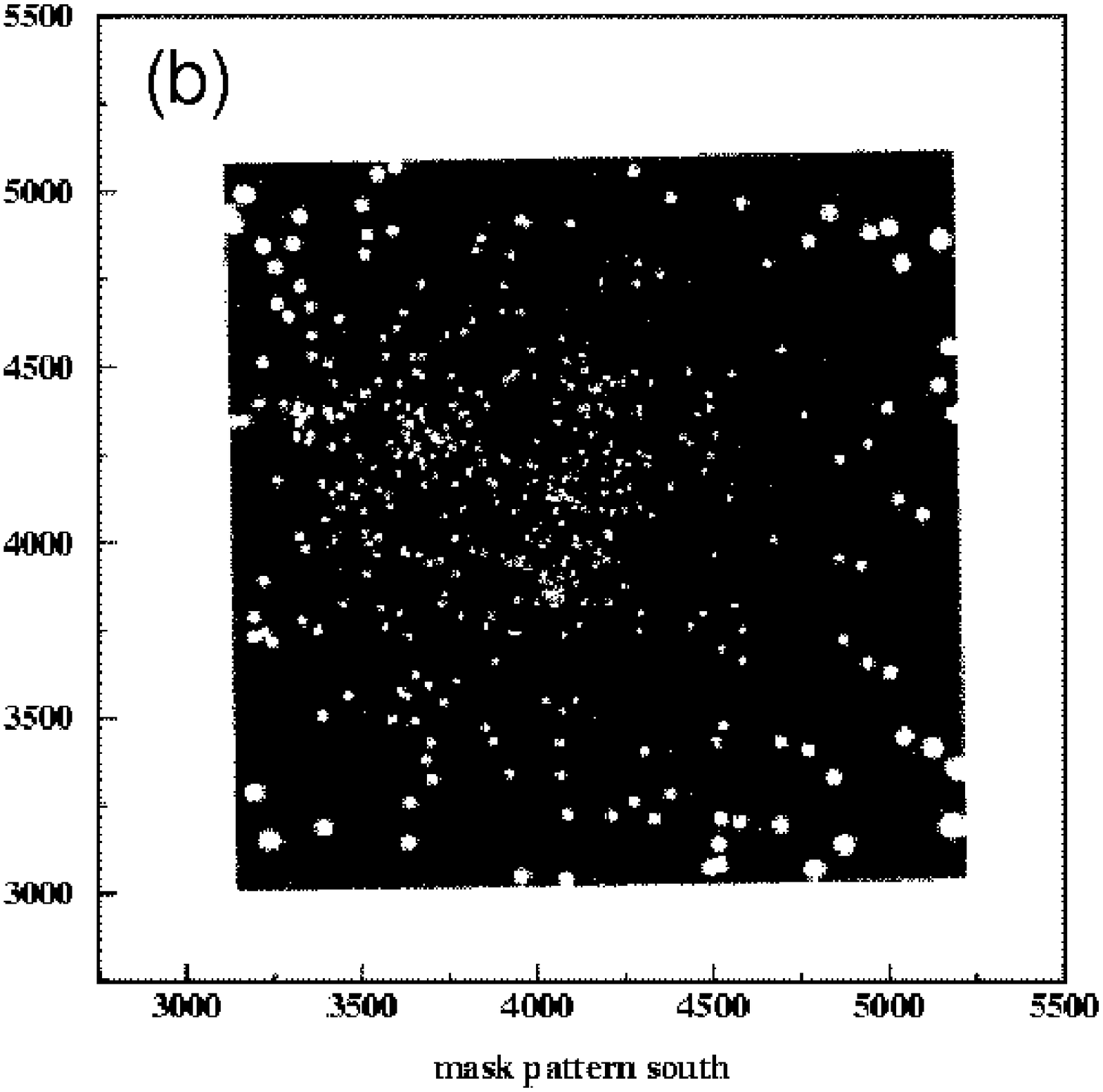}
\caption{Point-source mask patterns for the (a) north and (b) south 
         field observations, generated using the mask radius given by
         the ChaRT program.
         }
\label{fig:ngc6334:chandra:psext:mask}
\end{figure}

We have so far utilized signal extraction regions specified by the 
wavedetect routine. Although these regions are optimum for the source 
detection, they are usually too small to thoroughly remove photons from the point source.
Accordingly, we defined new circular regions around individual sources, 
based on the ``Chandra Ray Tracer''$\footnote{http://cxc.harvard.edu/chart/threads/index.html}$ (ChaRT).
Specifically, we calculated the encircled-photon fraction as a function 
of energy and
separation angle from the aimpoint (0, 1, 2, 3, 5, 7, 10, and 15 arcminites).
For each source, we looked up the results, using interpolation if necessary,
and chose a radius to include $\sim98$ \% of photons at the Al K$\alpha$-line 
energy (1.497 keV). 
Figure \ref{fig:ngc6334:chandra:psext:mask} shows point-source
mask patterns, obtained by summing up these larger regions around 
the individual point sources. In this procedure, 8\% and 7\% of the 
whole ACIS--I area have been excluded from the north and south field, 
respectively.

\section{Analysis of Extended X-ray Emission}
\label{sec:ana}

\subsection{Images}
\label{sec:ana:image}

\begin{figure}[htbp]
%%{fig/filth_0.5-2_n_cmap.eps}
%%{fig/filth_2-7_n_cmap.eps}
%%{fig/filth_0.5-2_s_cmap.eps}
%%{fig/filth_2-7_s_cmap.eps}
\centerline{
\includegraphics[height=7cm,angle=0,clip]
{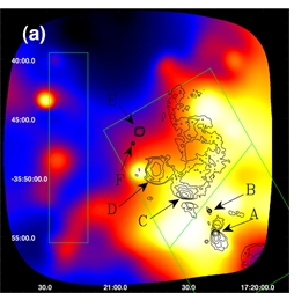}
\hspace*{0.5cm}
\includegraphics[height=7cm,angle=0,clip]
{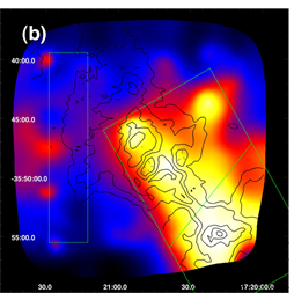}
}
\bigskip
\bigskip
\centerline{
\includegraphics[height=7cm,angle=0,clip]
{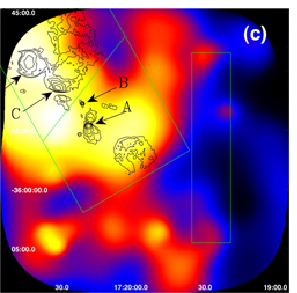}
\hspace*{0.5cm}
\includegraphics[height=7cm,angle=0,clip]
{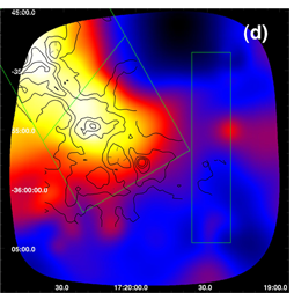}
}
%%\plottwo{f4a.eps}{f4b.eps}
%%\plottwo{f4c.eps}{f4d.eps}
\caption{Adaptively smoothed 
         images of the extended emission in the 0.5--2 keV (left)
         and 2--7 keV (right) bands, obtained by removing
         point sources using the mask patterns of figure
         \ref{fig:ngc6334:chandra:psext:mask}.
         The top and bottom panels correspond to the north and
         south observations respectively.
         They are corrected for the exposure
         and vignetting, but the background is not subtracted.
         Squares and trapezoids indicate regions
         utilized in the spectral analysis.
         Contours indicate the 18 cm radio continuum
        (\protect\citealt{Sarma2000}, left) and the CO $J$=2-1 maps
        (\protect\citealt{Kraemer1999a}, right).
         }
\label{fig:ngc6334:chandra:dmfilth}
\end{figure}

We applied the point-source masks (figure \ref{fig:ngc6334:chandra:psext:mask})
to the raw ACIS images to remove point sources, and then created images of the
residual emission using the CIAO tools {\bf dmfilth} and
{\bf csmooth}. The former fills holes in the image with
values interpolated from surrounding regions, and the
latter adaptively smoothes the filled image. We utilized 
default parameters (kernel sizes and significances) in
the {\bf csmooth} routine.
We corrected these images for the vignetting and exposure times.
Figure \ref{fig:ngc6334:chandra:dmfilth} shows the images
of the residual emission in two energy bands, obtained in this way.
To retain the best positional 
resolution available, we here show the two images separately 
instead of merging them together. In addition, 
the energy range is restricted to below 7 keV, 
to avoid the strong instrumental Ni K$\alpha$ line in the ACIS background.
                                                                               
As is clear from the procedure, the images shown in figure
\ref{fig:ngc6334:chandra:dmfilth} are thought to represent
the apparently extended emission noticed in figure
\ref{fig:ngc6334-chandra-csmimage}, although they
must be examined carefully against possible contribution
from fainter (and hence individually undetectable) point sources.
The overlapping region appears in somewhat different ways between
the north and south field images. This is due to the 
position-dependent angular resolution as described in \S 
\ref{sec:ps-extract}.

\subsection{Spectrum}
\label{sec:ana:spec}

\subsubsection{Region definition}
\label{sec:ana:spec:reg}

In order to analyze the spectrum of the extended emission,
we define in figure \ref{fig:ngc6334:chandra:dmfilth}
a rectangle of $\sim10^{\prime}\times18^{\prime}$,
elongated along the linearly aligned cores, and call
it ``extended emission region'', or EER for short.
Using the equilateral line from the two aimpoints, we 
further subdivide this EER into two trapezoids.
When analyzing the north-eastern and south-western
trapezoids, we use only the north and south field observation
data, respectively. This ensures that the overlapping region
is always placed within $\sim8^\prime$ of either of the two
aimpoints.
 
Because NGC 6334 is located on the Galactic plane, it is
contaminated by the diffuse emission along the Galactic ridge
(e.g., \citealt{Kaneda1997,Valinia1998,Ebisawa2001}).
In order to remove this unwanted component, we need to subtract
local background rather than those from blank skies.
We have therefore chosen two square background regions where the extended
emission is relatively weak (see figure \ref{fig:ngc6334:chandra:dmfilth}).
Hereafter we collectively call them ``background region'' or BR for short,
and utilize the events summed over them as our background.
The total area of the EER is 2.80/2.55$\times10^6$ pixel$^2$ 
before/after excluding the areas around point sources.
These areas correspond to 191/174 arcmin$^2$ or 47/43 pc$^2$ 
in 1.7 kpc. Thus, the excluded area is $\sim$10 \%. Similary,
those of the BR are 1.57/1.50$\times10^6$ pixel$^2$.

We then extracted spectra from the EER. The weighted arf and rmf files
were calculated using the CIAO programs {\bf mkwarf} and {\bf mkwrmf},
respectively, which take account of position-dependent detector responses
such as vignetting and energy resolution. The {\bf apply\_acisabs}
script was utilized when creating arf files, to correct them for 
the recent decrease in the ACIS quantum efficiency. We thus obtained two 
sets of spectrum, arf, and rmf files corresponding to the two observations, 
and then added them together considering the difference in their exposure 
times. In this way, we have acquired the combined spectrum
of the EER. We similarly derived a background spectrum from the BR.
Below, we utilize them after normalizing to the detector area (pixel$^2$), 
excluding the removed point sources. Since the effective area (cm$^2$) 
and also the instrumental background varies from place to place within 
ACIS--I, we must consider the positional difference of the 
sky and non X-ray backgrounds between the EER and BR. 
In the next section, we hence examine the excess emission 
considering these uncertainties.  

\subsubsection{Excess emission}
\label{sec:ana:spec:bgd}

Figure \ref{fig:ngc6334:chandra:diff:spec:bgcomp} (a) compares the spectrum
of the EER with that derived from the BR, both normalized to the same 
detector area. The former is higher than the latter by a 
factor of two in the 2--5 keV range, while they agree in energies below 
0.7 keV and above 7 keV where the instrumental background is dominant. 
The 0.5--7 keV raw counts of the EER and BR spectra are 22,500$\pm$150 and 
14,600$\pm$160, respectively, yielding the excess counts of 7,900$\pm$220. 
Thus, the excess emission is statistically quite significant.

To further confirm the significance of the extended emission, 
we investigated the systematic uncertainties
in the background. First, we compared the 5--10 keV count rates of two 
blank-sky spectra, extracted from the EER and BR locations of a blank-sky observation available to us, 
and examined the instrumental background for any position dependence.
We found negligible ($\sim$1 \%) difference between the two 
detector regions. Second, to examine how much 
the vignetting effect affects the amount of the sky background 
(Galactic ridge emission and cosmic X-ray background), 
we compared effective areas of the EER and BR specified
by their response files. The BR area was found to be smaller than
the EER area by $\sim$5\% at $<$5 keV, and by $\sim15$\% at 7 keV;
the resultant increase of the sky background is only 4\% of the extended 
emission. Thus, the excess counts in the EER can be concluded to be significant
even considering systematic errors.

\subsubsection{Comparison with summed point sources}
\label{sec:ana:spec:ps}

\begin{figure}[htbp]
%%{fig/diffuse_v5_sn_bgcomp.eps}
%%{fig/diffuse_v5_sn_src_v5_sn_comp.eps}
%%{fig/diffuse_v5_sn_src_v5_sn_divcomp.eps}
\centerline{
\includegraphics[height=8cm,angle=-90,clip]
{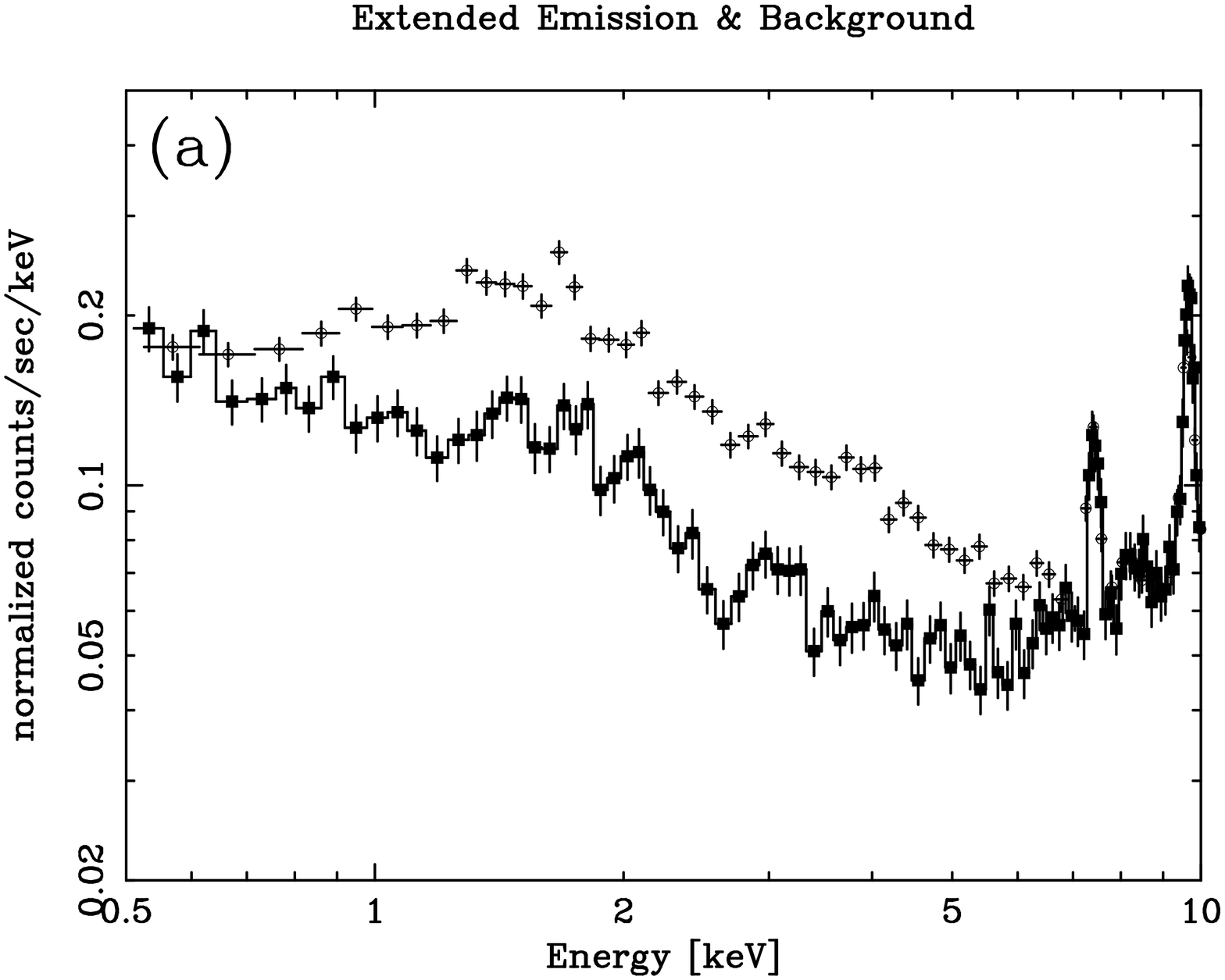}
\hspace*{0.5cm}
\includegraphics[height=8cm,angle=-90,clip]
{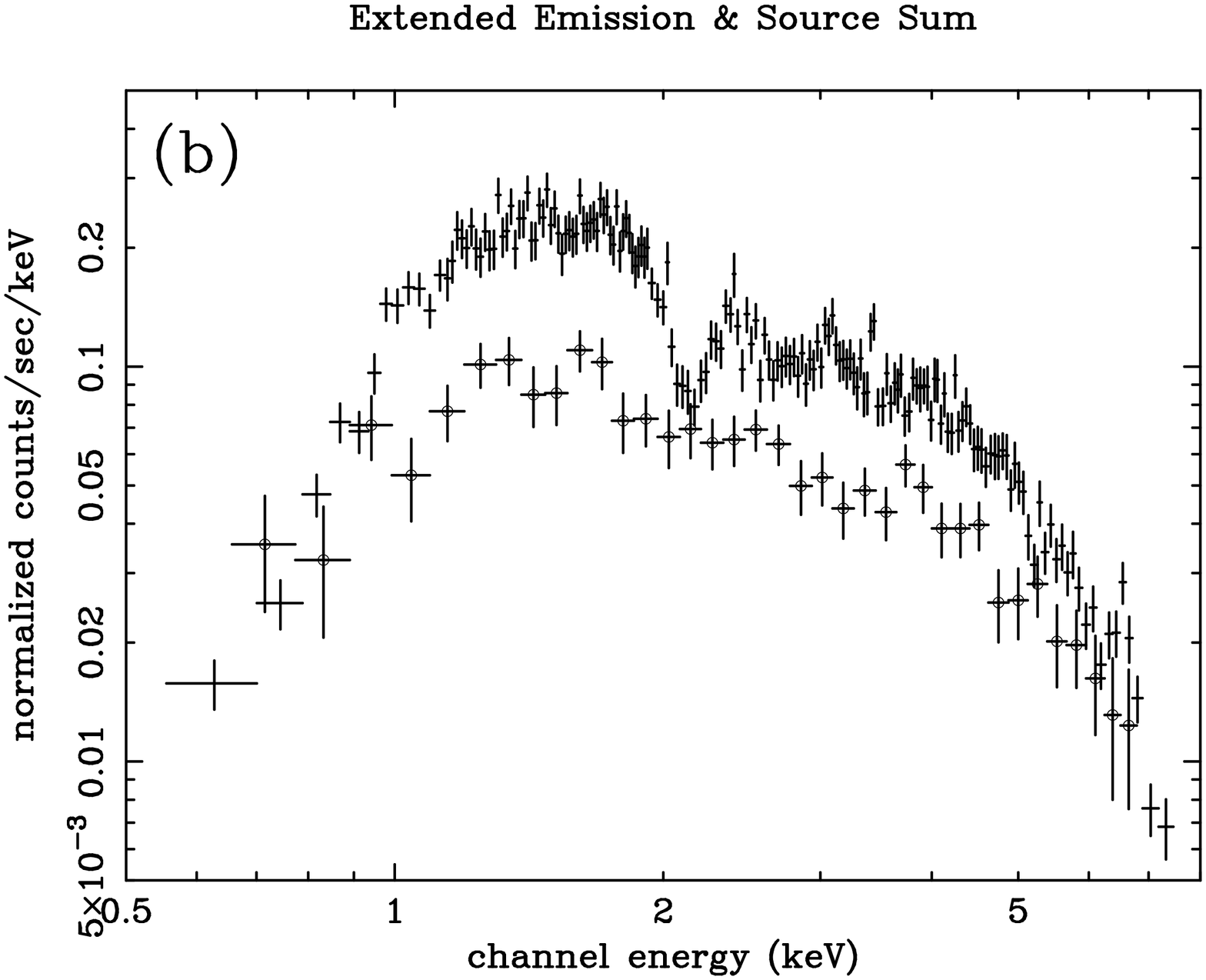}
}
\bigskip
\centerline{
\includegraphics[height=7cm,angle=0,clip]
{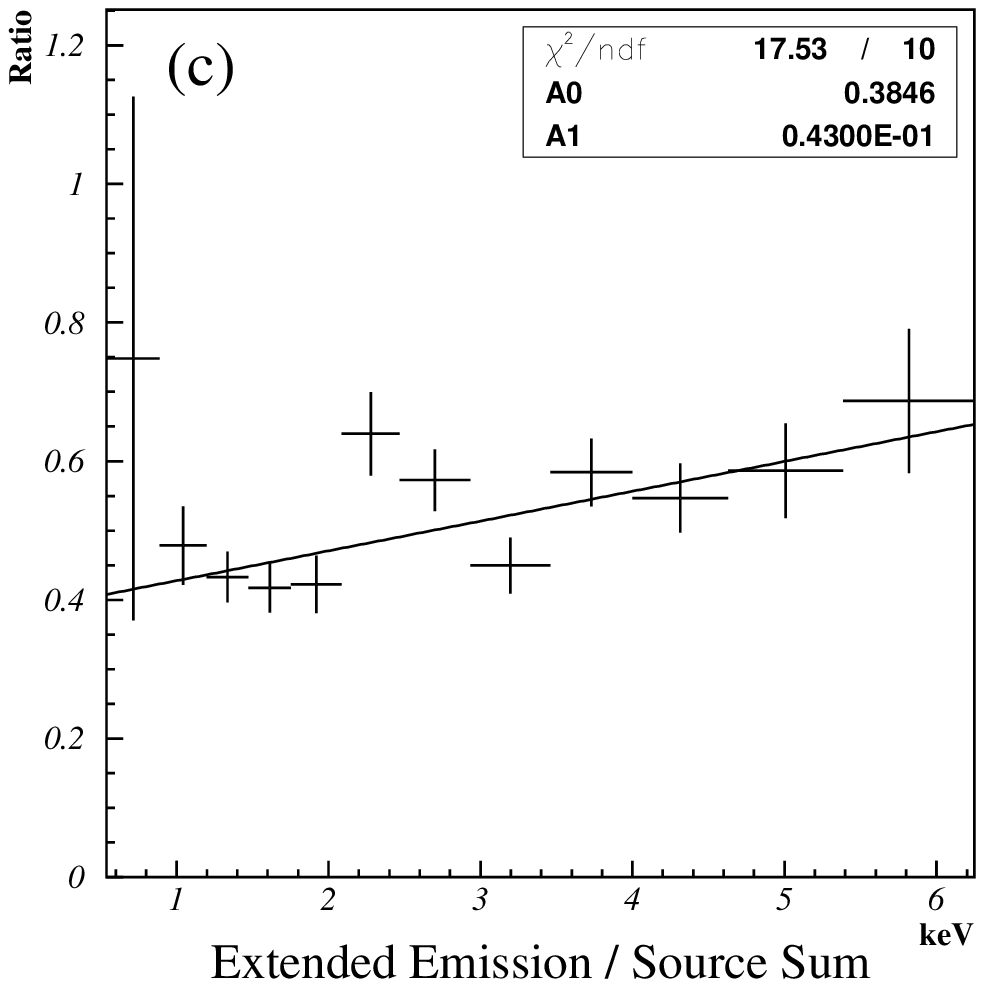}
}
%%\plottwo{f5a.eps}{f5b.eps}
%%\plotone{f5c.eps}
\caption{(a) The ACIS spectrum of the EER (circle) 
             compared with that of the BR (filled square).
         (b) The background subtracted spectrum of the EER (circle) 
	     and that summed over 548 point sources (cross).
	 (c) Ratio of the background-subtracted EER spectrum to the
             source sum spectrum, as a function of X-ray energy.
             The best-fit first order polynomial is shown as a solid line.
         }
\label{fig:ngc6334:chandra:diff:spec:bgcomp}
\label{fig:ngc6334:chandra:diff:spec:srccomp1}
\label{fig:ngc6334:chandra:diff:spec:srccomp2}
\end{figure}

In order to characterize the spectrum of the background-subtracted
EER emission, we compare it in figure 
\ref{fig:ngc6334:chandra:diff:spec:srccomp1} (b)
with the ACIS spectrum summed over the 
548 point sources detected within the EER. 
This figure immediately reveals several important features of the
extended emission.
Firstly, it is nearly half as luminous as the emission summed 
over all the detected point sources. Secondly, the extended emission 
exhibits a harder spectral continuum in the 2--7 keV range.
Third, no significant emission line is observed in the EER spectrum,
in contrast to the point-source spectrum which shows several emission
lines from elements such as S, Ar, and Fe. (Incidentally, the simultaneous
presence of these lines is not surprising, since we have added sources
with various temperatures).
In order to better visualize this, we show the ratio of the two spectra
in figure \ref{fig:ngc6334:chandra:diff:spec:srccomp2} (c). By fitting the
spectral ratios with a linear function of energy,
given as ($A_0+A_1\times$Energy), we obtained
$A_0=0.38\pm0.03$ and $A_1=0.043\pm0.012$ where errors are 1$\sigma$.
The two spectra significantly differs each other.

\subsubsection{Flux and luminosity}
\label{sec:ana:spec:fit}
 
\begin{figure}[htbp]
%%{fig/diffuse_v5_sn_po.eps}
%%{fig/src_v5_sn.eps}
%\plottwo{f6a.eps}{f6b.eps}
\centerline{
\includegraphics[height=8cm,angle=-90,clip]
{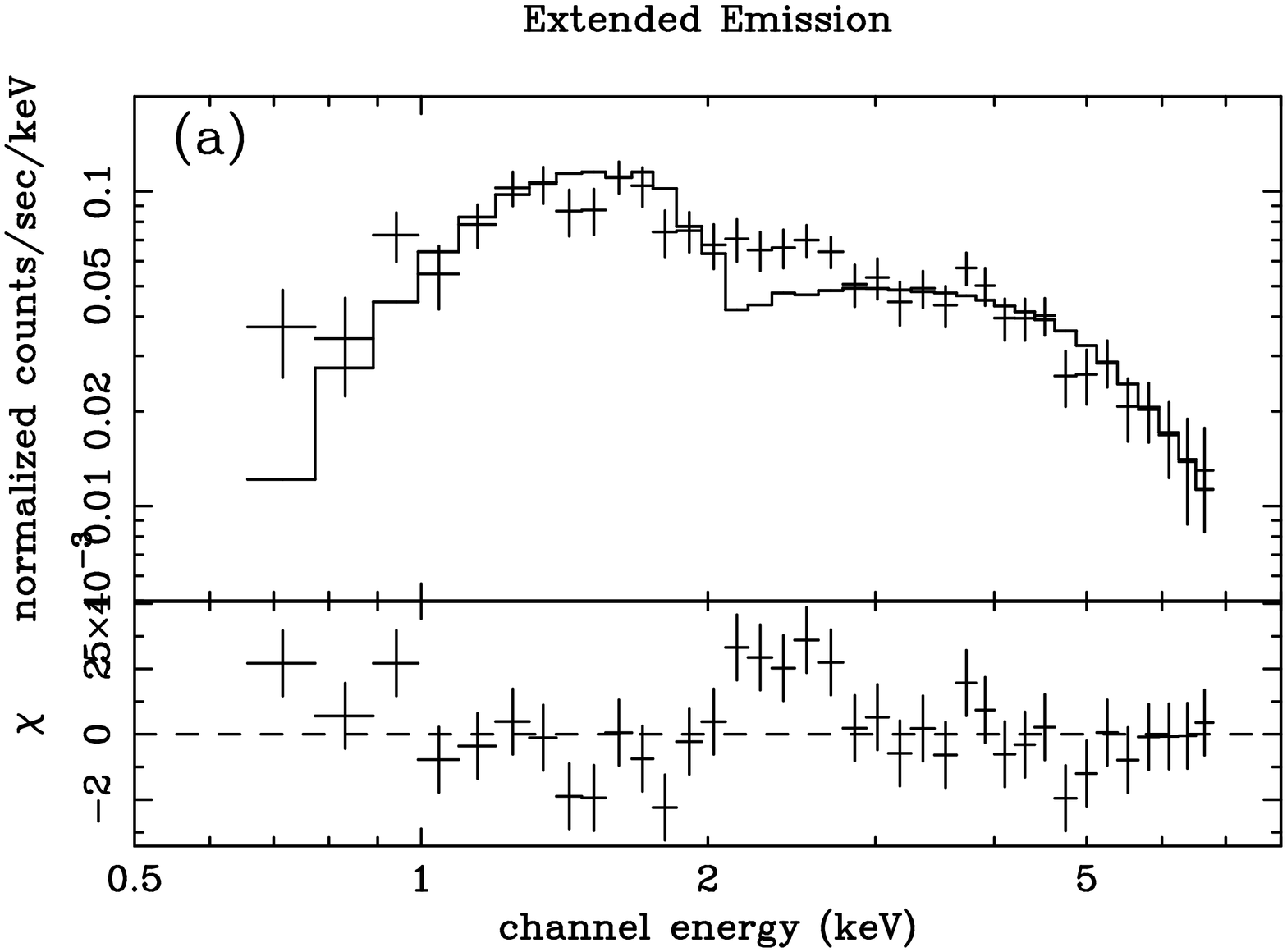}
\hspace*{0.5cm}
\includegraphics[height=8cm,angle=-90,clip]
{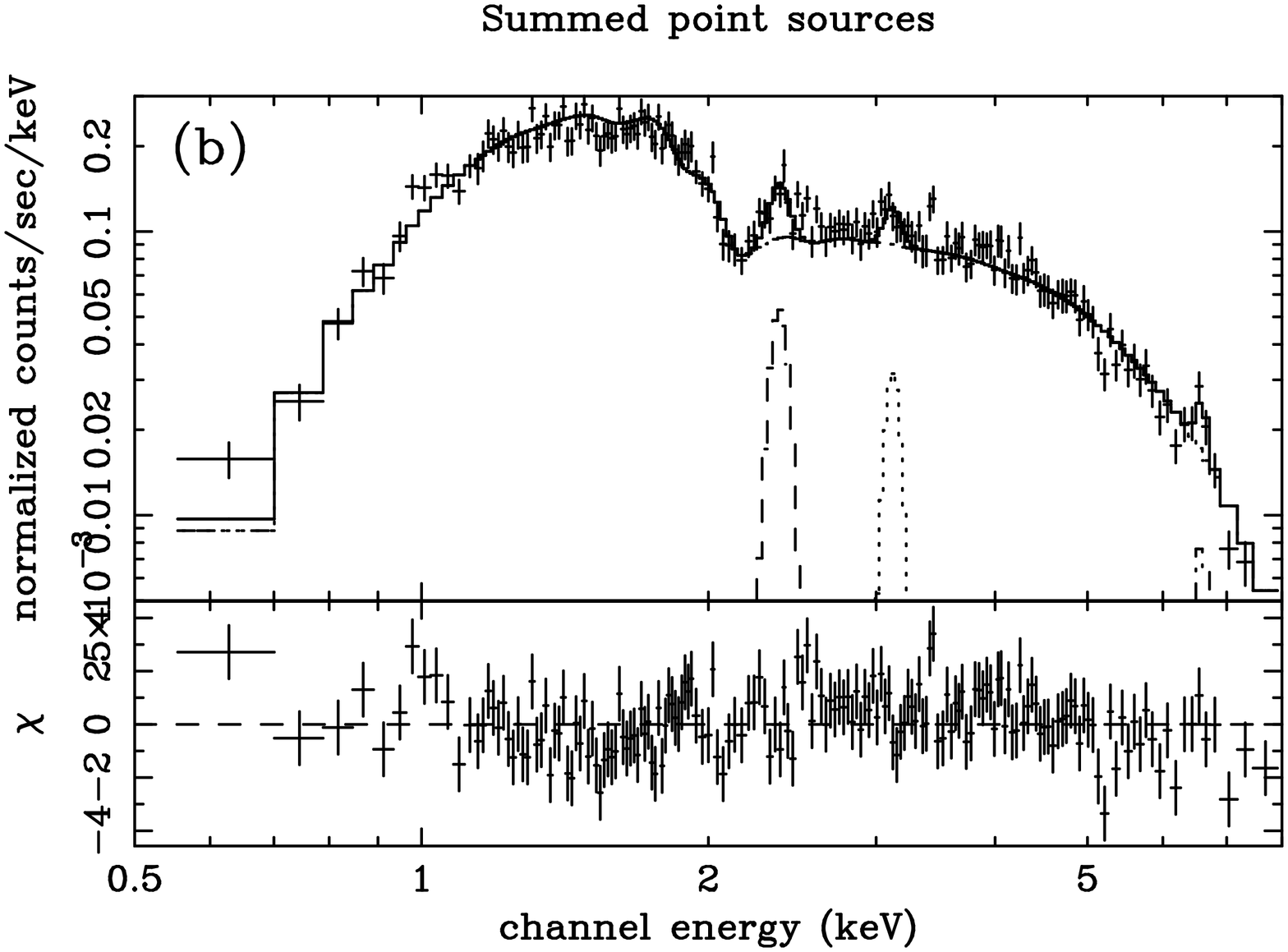}
}
\caption{(a) The ACIS--I spectrum of the EER, compared 
         with the best-fit power-law model (solid line). See table
         \ref{tbl:ngc6334:chandra:diff:spec:param1} for the obtained
         parameters.
         (b) The same as panel (a) but for the summed point sources.         
         The solid line indicates the best-fit power-law plus three 
         Gaussian model. See table \ref{tbl:ngc6334:chandra:diff:spec:param2} 
         for the parameters. 
         }
\label{fig:ngc6334:chandra:diff:spec:fit1}
\label{fig:ngc6334:chandra:diff:spec:fit2}
\end{figure}

We conducted spectral fitting to the EER spectrum, to quantify 
its basic properties such as the X-ray flux and luminosity.
We here employed a simple power-law model with an
interstellar absorption. Here and hereafter, all quoted errors in the
spectral fitting refer to 90 \% confidence levels unless otherwise
stated. Table \ref{tbl:ngc6334:chandra:diff:spec:param1} lists
the obtained parameters, and 
figure \ref{fig:ngc6334:chandra:diff:spec:fit1} (a)
shows the fitting results. The fit was not acceptable with 
$\chi^2/\nu\sim2$ because of an excess around 2--3 keV, although it 
contributes only $\sim$5\% in the 0.5--7 keV flux. The 0.5--8 keV 
flux implied by the best-fit model is 5.6$\times10^{-12}$ \flux 
before removing the absorption, and the absorption-corrected 
0.5--8 keV luminosity reaches 2$\times10^{33}$ \ergs.

For comparison, we quantified the summed point-source 
spectrum in terms of a power-law model. 
Additionally, three Gaussians were utilized to reproduce 
the H-like Si K$_\beta$ (line1), He-like Ar K$_\alpha$ (line2), 
and a complex of neutral and He-like Fe K$_\alpha$ (line3) lines,
seen in the spectrum. We obtained results as shown in table 
\ref{tbl:ngc6334:chandra:diff:spec:param2} 
and figure \ref{fig:ngc6334:chandra:diff:spec:fit2} (b);
the relatively hard continuum ($\Gamma\sim1.2$) is considered to arise from 
a superposition of plasma emission absorbed with different column densities.

We also consider to what extent the EER spectrum is contaminated 
by photons which escaped from the summed point sources. 
We then multiplied the above model for the summed point sources by the 
escape-fraction curve estimated by the ChaRT program, and re-fitted the 
EER spectrum considering this contribution. This has little changed the 
results, except an $\sim$10\% decrease of the EER flux. 
Thus, the contamination of the detected point sources is small.

\subsection{Luminosity Function}
\label{sec:ana:lumi}

The {\it Chandra} data of NGC 6334 thus reveal an apparently
extended hard X-ray emission, which remains significant 
after removing the detected point sources. Although this 
emission could be truly diffuse, it could alternatively be formed 
by a collection of faint point sources which are individually 
undetectable. In order to examine this issue,
we utilize the luminosity function of the detected point sources.
Here we define it by a cumulative column number density (pc$^{-2}$)
of those sources of which the absorption-uncorrected 0.5--8 keV
luminosity is higher than a specified value, $L$ (\ergs). 

\begin{figure}[htbp]
%%{fig/lognlogs_lt_v5.eps}
%%\plotone{f7.eps}
\centerline{
\includegraphics[height=15cm,angle=-90,clip]
{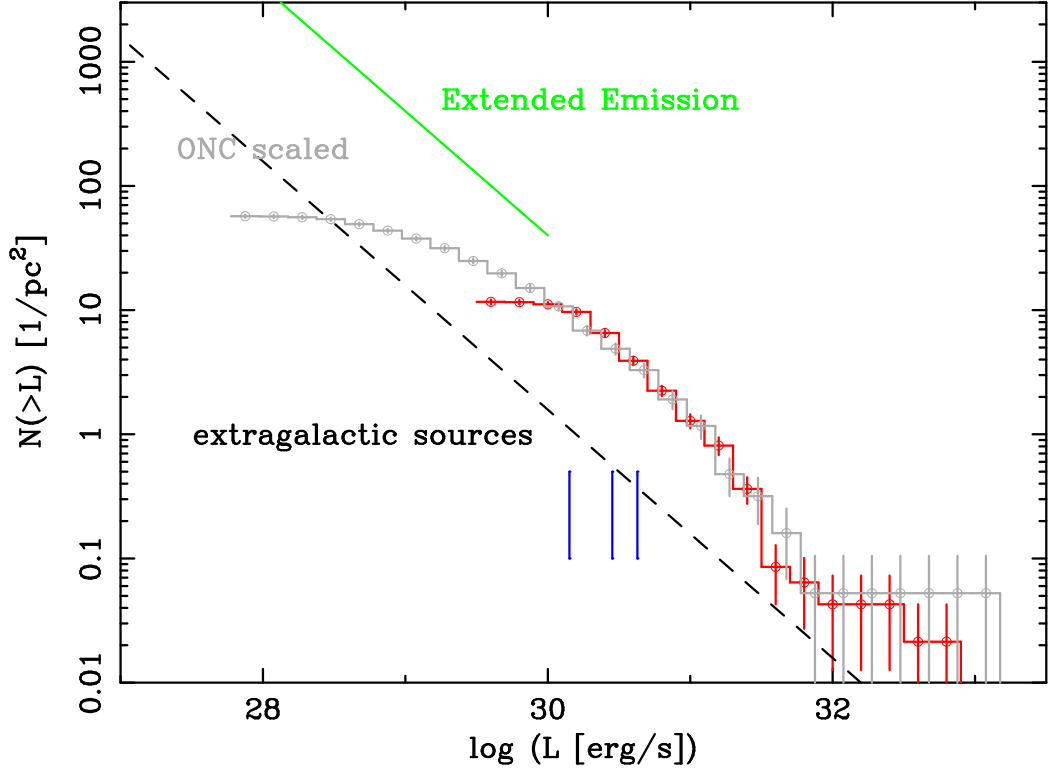}
}
\caption{The luminosity function of the point sources within the EER
         (red), compared with that of ONC (gray)
        taken from \protect \cite{Feigelson2002a}.
        The vertical axis is the column density of the source number, while the
        horizontal axis absorption-uncorrected 0.5--8 keV luminosity.
        The ONC curve is scaled considering absorption and source
        density (see text). Errors are 1$\sigma$ Poisson.
        The green line indicates the number of putative
        point sources at a given flux that can account for the
        observed total surface brightness of the extended emission.
        Three vertical lines near the bottom are the typical X-ray fluxes of 
        point
        sources yielding 10, 20, and 30 netcounts in a 40 ksec observation.
        The dashed line indicates the expected number of extragalactic
        sources.
        }
\label{fig:ngc6334:chandra:diff:spec:lognlogs}
\end{figure}

Figure \ref{fig:ngc6334:chandra:diff:spec:lognlogs} shows the
luminosity function of the 548 point sources in the EER;
we derived $L$ of the bright sources ($\geq$30 netcounts) 
by spectral fitting. We fitted each spectrum with a plasma emission
code called {\bf apec} (astrophysical plasma emission code)$\footnote{http://hea-www.harvard.edu/APEC/}$
or power-law model, whichever gives a better fit to the data.
Details will be presented in Ezoe et al. (in preparation).
For the other faint sources ($<$ 30 netcounts), we utilized the 
count-to-flux conversion factor derived from the summed point sources
(figure \ref{fig:ngc6334:chandra:diff:spec:fit2} b);
$\sim2\times10^{-11}$ \flux (netcounts/s)$^{-1}$.

In figure \ref{fig:ngc6334:chandra:diff:spec:lognlogs}, 
the source number density increases toward lower luminosities, and 
saturates below $\sim10$ netcounts, corresponding to the completeness limit
estimated in \S \ref{sec:ps-extract}.
Specifically, the 
completeness limit becomes about 30 netcounts/(40 ksec), or 
$5\times10^{30}$ \ergs in terms of absorption-uncorrected 0.5--8 keV 
luminosity. Then, in order to complement the luminosity function below
this limit, we have incorporated {\it Chandra} results on the Orion Nebula 
Cluster (ONC) by \citet{Feigelson2002a}, which also utilize an
absorption-uncorrected luminosity function. This is a representative 
and very nearby ($D=450$pc) MSFR, and the completeness limit of \citet{Feigelson2002a},  
in terms of the 0.5--8 keV luminosity, 
is lower by a factor of 20 than the present limit for NGC 6334.
Furthermore, the stellar mass function of NGC 6334 and that of the Orion 
Nebula are known to have consistent shapes within errors, as indicated
by NIR observations \citep{Straw1989a}. 
Another justification for using the ONC result is provided by the fact
that nearby young stellar clusters have very similar low-mass stellar 
populations, and hence similar X-ray luminosity functions toward the 
low-luminosity end \citep{Feigelson2004}.
Therefore, the ONC result is ideal for interpolating the luminosity 
function of NGC 6334 below our detection limit.

To be more accurate, we would have to compare the two luminosity functions
after correcting for the absorption. Considering this, we scaled that of 
ONC by a factor of 1.3 toward lower luminosities, because typical column 
densities in ONC and NGC 6334 are $\sim 0.5\times10^{22}$ and 
$\sim 1.3\times10^{22}$ cm$^{-2}$, respectively 
(\citealt{Feigelson2002a,Dickey1990}), 
and hence a typical YSO with $kT\sim3$ keV would be subject to a larger 
correction for absorption if it is in NGC 6334 than in ONC. 
In the vertical 
direction, we arbitrarily scaled the luminosity function of ONC by a 
factor of 0.3, so that it coincides with that of NGC 6334 at the 
luminosity of the completeness limit.
This scaling factor, $<1$, is consistent with the fact that
the EER contains not only star-forming cores but also molecular
cloud envelopes; the EER contains $\sim$7 star-forming cores in
47.1 pc$^2$, while ONC has 1 in 4.9 pc$^2$. Thus, the stellar
density must be higher in ONC. After these shifts, the two
luminosity functions coincide with each other very well within Poisson
errors, in luminosities above $5\times10^{30}$ \ergs. This reconfirms that 
the X-ray source population is similar between these two objects, and that 
the estimation of the completeness limit is reliable.

We estimated the expected flux of unresolved sources by integrating 
the rescaled ONC luminosity function, from the completeness limit down 
to the end, $\sim10^{28}$ \ergs. From this, we subtracted the number 
of point sources actually detected in our observations at luminosities 
below the completeness limit. Then, the estimated unresolved sources 
have turned out to be 2500 in number, and $\sim0.5\times10^{31}$ erg 
s$^{-1}$ pc$^{-2}$ in surface brightness. This is only 12 \% of the 
value needed to account for the surface brightness of the extended 
emission. We further extrapolated the function into 0 \ergs assuming a 
slope of 0.4, taken from that of ONC in the 10$^{29-30}$ \ergs range;
we found that it reaches at most $\sim$20 \%, and hence falls still short 
of the extended emission.

In order to estimate the expected flux from the background extragalactic 
sources, we utilized the source density from \cite{Giacconi2001}. 
We converted the flux into the luminosity, and the surface number density 
of sources into their column density, both assuming that all of them are 
at the distance of NGC 6334. The flux estimated in this way 
is negligible (3\% in $10^{27}-10^{28}$ \ergs). Furthermore, the strong 
concentration of the extended emission on the EER cannot be explained by 
background objects. 

In conclusion, a large part ($\simgt80$\%) of the extended emission is 
suggested to be truly diffuse in nature. The spectral difference between
the diffuse emission and the point sources (\S \ref{sec:ana:spec:ps}, fig \ref{fig:ngc6334:chandra:diff:spec:srccomp2} c) independently supports this conclusion.

\section{Region-by-region Spectral Analysis}
\label{sec:region-by-region}

Although the apparently flat continuum of the EER, represented by a 
power-law index of 0.9, is suggestive of non-thermal emission, 
such a flat continuum could alternatively be produced by a 
superposition of thermal components with different absorptions. 
Accordingly, in this section, we divide the EER into finer regions, 
and study the spectrum and X-ray absorption as a function of the position.

\subsection{Region selection}
\label{sec:region-by-region:region}

\begin{figure}[htbp]
%%{fig/filth_0.5-2_n.eps}
%%{fig/filth_2-7_n.eps}
%%{fig/filth_0.5-2_s.eps}
%%{fig/filth_2-7_s.eps}
\plottwo{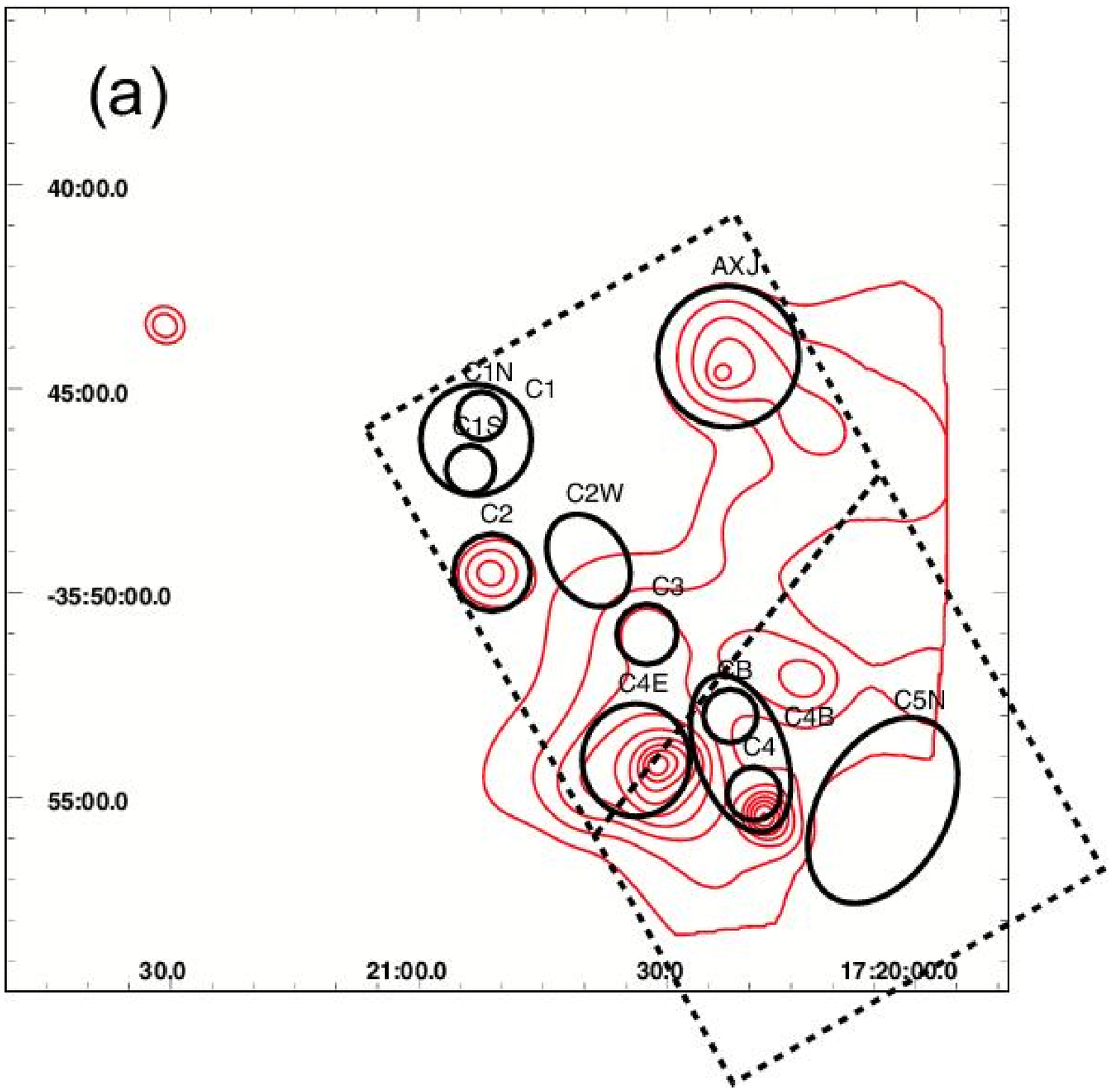}{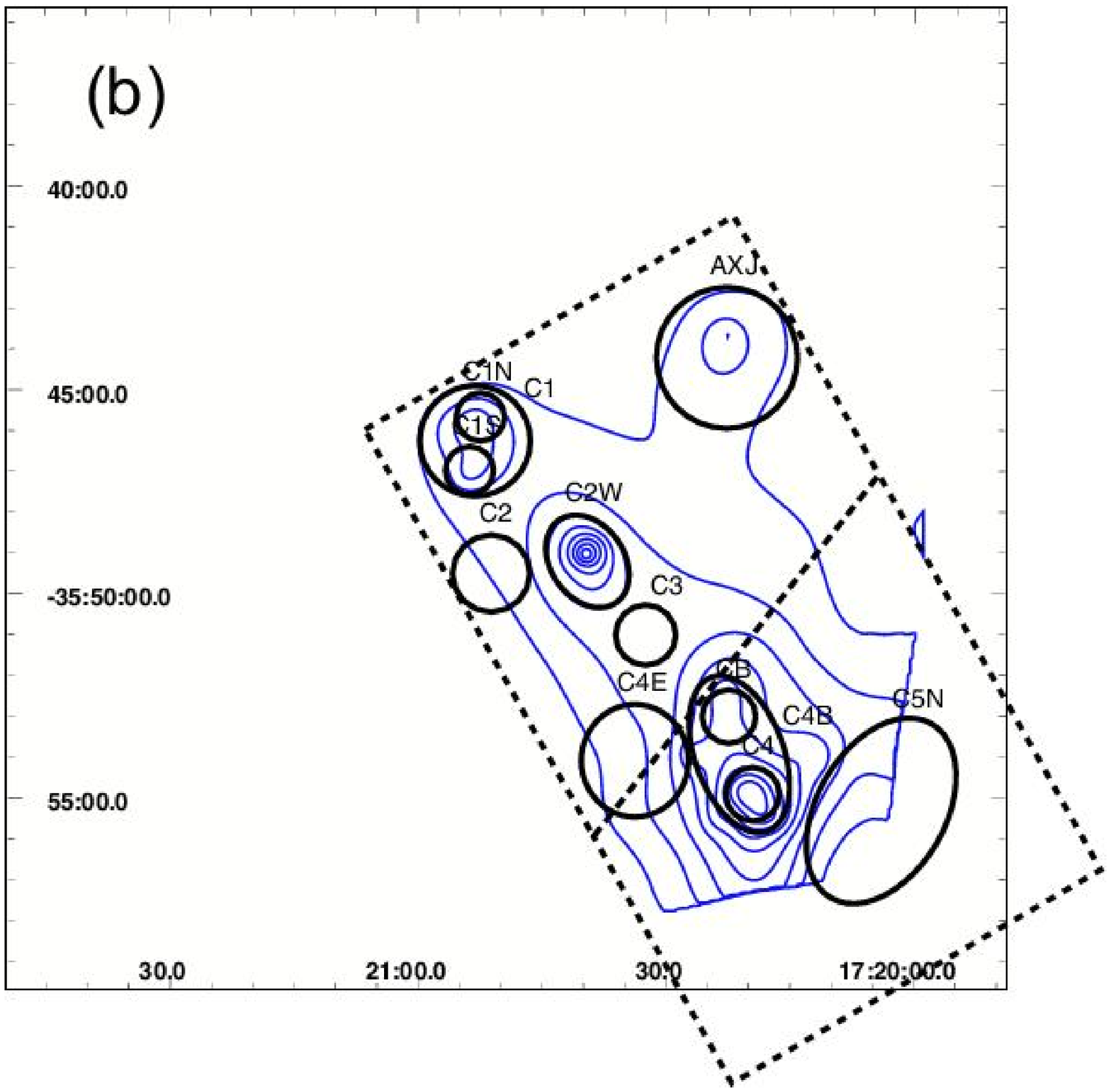}
\plottwo{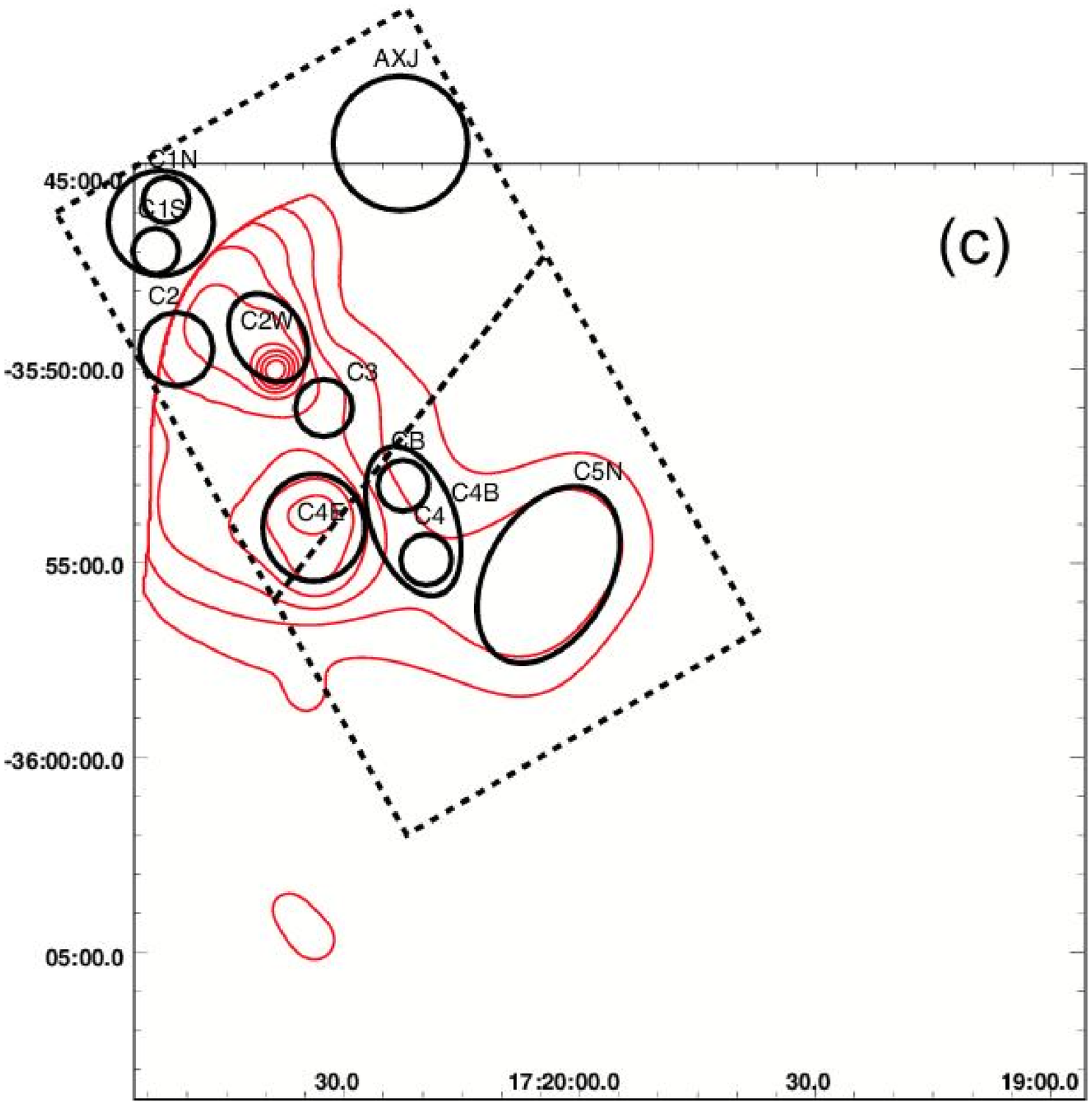}{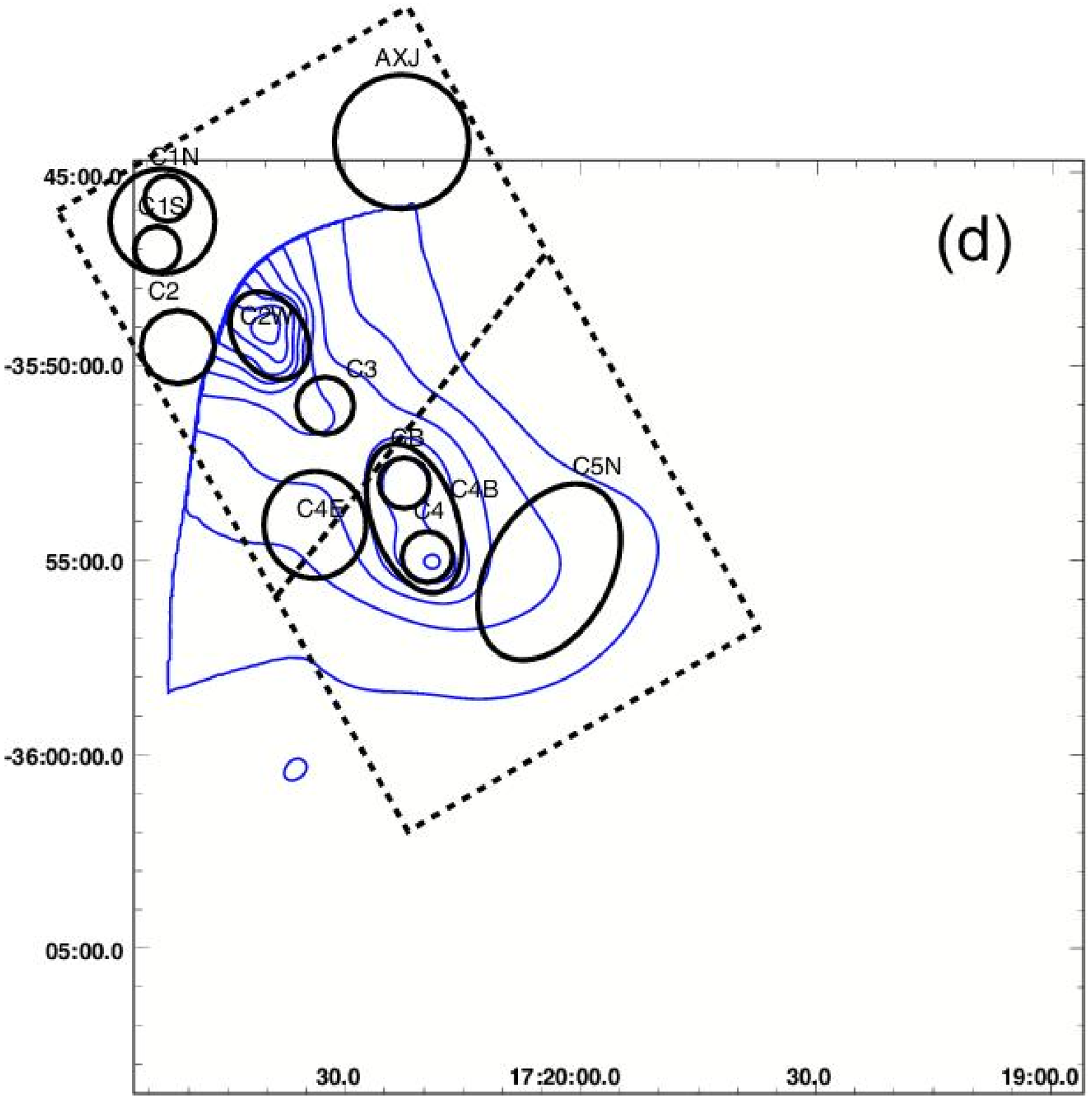}
\caption{The same as figure \ref{fig:ngc6334:chandra:dmfilth}
         but plotted with contours in logarithmic scales.
         Panels (a) and (c) show the soft band images, while (b) and (d)
         those in the hard band. Panels (a) and (b) represent the north
         field while (c) and (d) the south field.
         Solid circles and ellipses indicate the selected regions
         to be utilized in the spectral analysis. Two dashed trapezoids 
         represent the EER.
         }
\label{fig:ngc6334:chandra:dmfilth2}
\end{figure}

To conduct the spatially-resolved spectroscopy, we have reproduced
in figure \ref{fig:ngc6334:chandra:dmfilth2}
the two-band images of figure \ref{fig:ngc6334:chandra:dmfilth}, 
but utilizing this time
logarithmic contour representations. The contours are separated by
a factor of 1.1 and 1.2 in the background-inclusive brightness in the soft 
and hard band images, respectively, while the lowest contour represents
$\sim$1.3 times the average BBR value in both bands. In reference to
these two-band images, let us define characteristic regions, particularly
bright clumps, to be utilized in the subsequent analysis.

Figure \ref{fig:ngc6334:chandra:dmfilth2} (a), namely
the soft X-ray image of the north field, reveals several 
bright clumps coincident with the FIR cores or the condensations of 
massive stars. Referring to this contour map, we then define 
three circular regions, named C2, C3, and AXJ, with C standing for ``FIR core''
and AXJ the known X-ray source AXJ 1720.4$-$3544 which is identified
with a B0.5e star by \cite{Matsuzaki1999a}.
The C3 region is less clear than the other two, but it is chosen so 
as to include the known massive ZAMS (O8) at its center. 

In the hard X-ray image of the north field 
(figure \ref{fig:ngc6334:chandra:dmfilth2} b), 
these three regions become less clear. On the other hand, a bright 
clump emerges at the position of the FIR core I(N) and I. 
Accordingly, a new region C1 is defined, together with two more sub-regions
in it, C1N and C1S to represent the two independent FIR 
cores therein. Also, a significant emission can be seen 
to the west of the FIR core II, coincident with a molecular cloud 
dark lane and a condensation of embedded stars. Hence, we define 
another region named C2W 
region. After all, we have defined 7 regions in the north field 
(C1, C1N, C1S, C2, C2W, C3, and AXJ). 

Similarly, using figure \ref{fig:ngc6334:chandra:dmfilth2} (c)
which is the soft X-ray map of the south field, 
we select two bright soft X-ray regions, C4E and C5N, toward 
the east and north directions of the core IV and V, respectively. 
In the hard X-ray map (figure \ref{fig:ngc6334:chandra:dmfilth2} d), 
a bright clump stand out at the FIR 
core IV and its north position, which are covered by regions
named C4 and CB, respectively. A larger region C4B is employed to sum up 
C4 and CB. Thus, in the south field, we 
have defined 5 regions (C4E, C5N, C4, CB and C4B). 
After all, the EER has been subdivided into 12 representative regions.

\subsection{Color-color diagram}
\label{ngc6334:chandra:diff:pos:hr1hr2}

\begin{figure}[htbp]
%%{fig/hr1hr2-2.eps}
\plotone{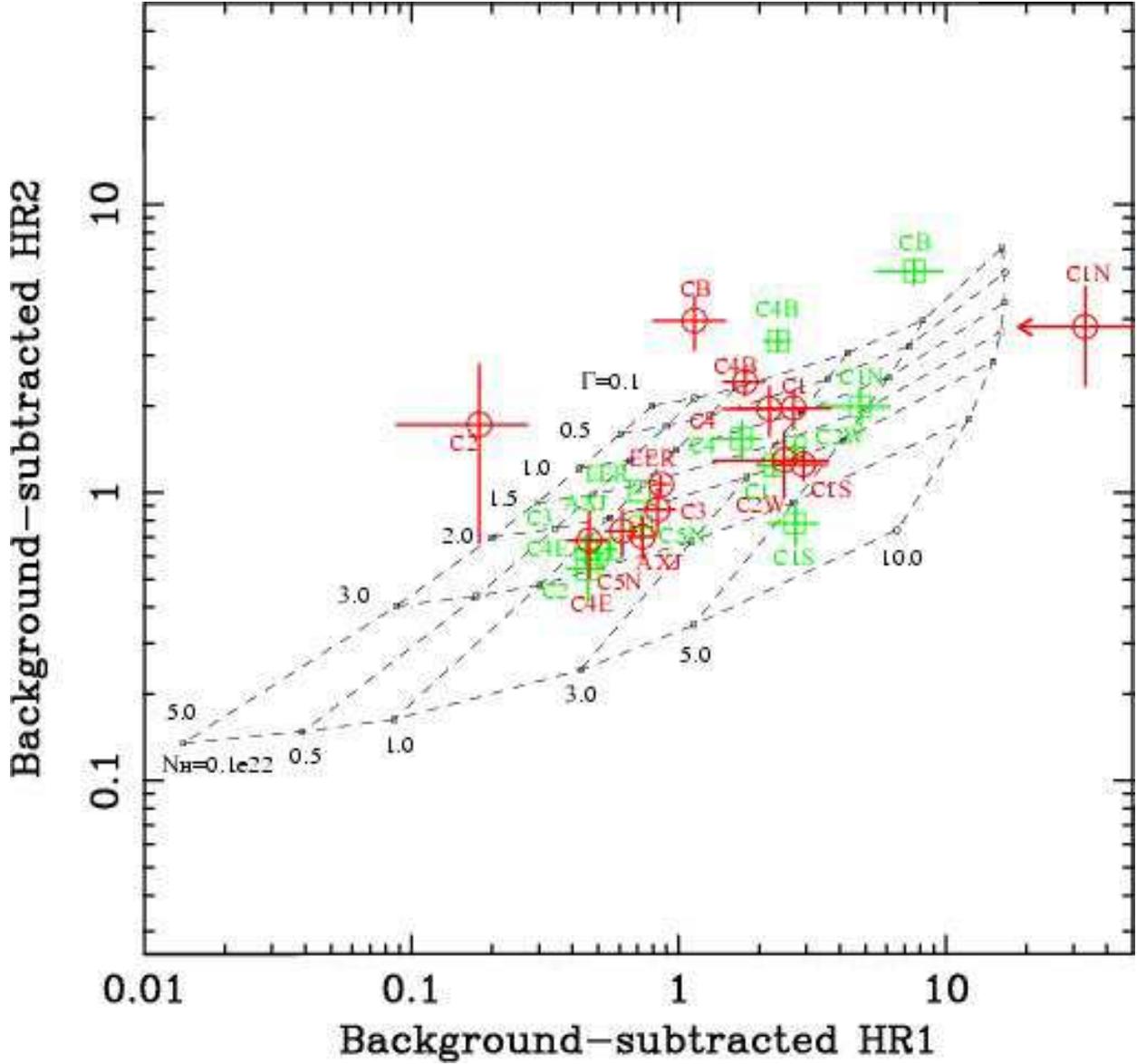}
\caption{The color-color diagram of the diffuse emission (red) and
         the summed point sources therein (green), measured in the 12 regions
         defined in figure \ref{fig:ngc6334:chandra:dmfilth2}.
         The error bars represent $\pm1\sigma$ statistical uncertainties.
         Dashed lines are predictions by a family of single power-law models,
         having different power-law indicies and suffering different 
         interstellar absorptions in a unit of 10$^{22}$ cm$^{-2}$.
         }
\label{fig:ngc6334:chandra:diff:pos:hr1hr2-2}
\end{figure}

To grasp the spectral properties of the 12 regions, 
we arranged them in figure \ref{fig:ngc6334:chandra:diff:pos:hr1hr2-2} 
on a color-color diagram. 
We divide the  0.5--7 keV band into three finer bands
($S$ 0.5--2.0, $M$ 2.0--3.5, and $H$ 3.5--7 keV), and created
two hardness ratio maps by calculating $HR_1=M/S$ and $HR_2=H/M$.
$HR_1$ is expected to mainly reflect a difference in an absorption
column density, while $HR_2$ is more sensitive to the intrinsic
continuum hardness.
The data points can be clearly subdivided into
two groups; (1) "soft" regions with $HR_1\lesssim1$
and $HR_2\lesssim1$, including AXJ, C2, C3, C4E and C5N;
and (2) "hard" regions with $HR_1\gtrsim1$ and $HR_2\gtrsim1$, 
including C1, C1N, C1S, C2W, C4, CB, and C4B.

For comparison, green points
in figure \ref{fig:ngc6334:chandra:diff:pos:hr1hr2-2}
show the hardness ratios of the summed point sources 
in the individual regions. The green data are again divided
into the two groups; the point sources in the soft regions 
exhibit smaller values of $HR_1$ than those in the hard ones. 
The $HR_2$ values are also different between the two groups, 
especially in CB and C4B, although  these two are contaminated 
by the bright background AGN (NGC 6334 B). 
Importantly, most of these summed point sources have 
similar or even higher values of $HR_1$ than the diffuse 
emission in the same region, suggesting that the latter 
suffers less absorption. 

\subsection{Spectral fitting for summed point sources}
\label{ngc6334:chandra:diff:pos:fitting:src}

Before analyzing the diffuse emission spectra, we analyze spectra
of the summed point sources with two 
aims in mind. One is to estimate their absorption column densities, which 
serve as a measure of absorption affecting the diffuse emissions. The other 
is to utilize their spectral shapes to estimate the effects of those photons
which escape from the point sources into the diffuse emission.

\begin{figure}[htbp]
%%{fig/nhktzsx.eps}
\plotone{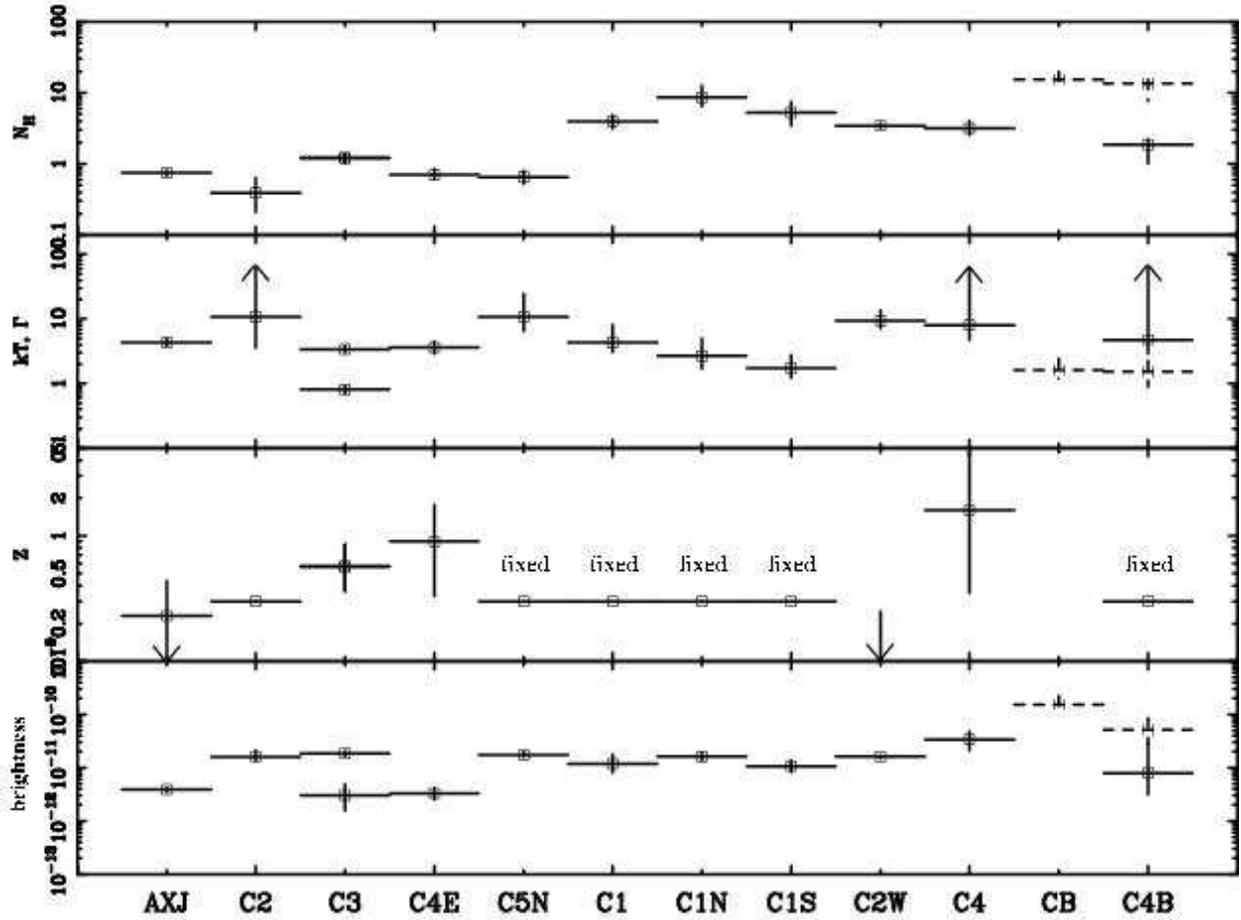}
\caption{Results of the spectral fitting to the summed point sources in 
         the 12 diffuse emission regions
         with a single temperature (plus Gaussian) model, 
         or a power-law model (dashed), or a combination of them.
         From top to bottom panels, plotted are the best-fit values
         of the absorption column density in 10$^{22}$ cm$^{-2}$, 
         the temperature in keV or photon index, 
	 the abundance in solar units, and 
         the uncorrected 0.5--8 keV 
         X-ray surface brightness in \flux pc$^{-2}$.
         }        
\label{fig:ngc6334:chandra:diff:spec:srcsum:summary}
\end{figure}

In the same way as in \S \ref{sec:ana:spec:ps},
we prepared the spectrum of the 
summed point sources in each region, and fitted it with simple models. 
The spectra in 8 out of the 12 regions have been represented successfully 
by a single temperature thin-thermal model. The C2W and C3 region 
required an additional narrow Gaussian and a second thermal plasma model
of a rather low temperature, respectively. 
For the remaining two regions (CB and C4B) which involve the bright AGN 
(NGC 6334 B), a power-law model and a single temperature plus 
power-law model, respectively, gave acceptable fits. 
All the 12 spectra have been successfully reproduced in this way, yielding
the results in figure \ref{fig:ngc6334:chandra:diff:spec:srcsum:summary}. 

The obtained absorption column density differs clearly 
between the soft and hard regions, confirming the inference 
from the color-color diagram. Specifically, the soft regions 
have column densities of $(0.5\sim1)\times10^{22}$ cm$^{-2}$, 
while the hard ones have $(2\sim10)\times10^{22}$ cm$^{-2}$. 
These values are consistent (within a factor of $\sim3$) with
those values estimated in other wavelength ranges,
 such as the radio CO line 
\citep{Kraemer1999a} and NIR extinctions \citep{Straw1989a}. 
Furthermore, every spectrum except those from 
three regions (C3, CB and C4B) has been successfully reproduced 
by a single temperature plasma model absorbed by a single column density.
This ensures that each of the 12 regions has a well-defined value 
of $N_{\rm H}$. The derived temperatures are moderately high, i.e., 
several keV, in agreement with the typical X-ray temperature of YSOs. 
Therefore, the detected point sources can be mostly understood as YSOs
suffering from region-dependent absorptions.

\subsection{Spectral fitting of the diffuse emission}
\label{ngc6334:chandra:diff:pos:fitting}

We then conducted model fitting to the diffuse emission spectrum
in each region. 
In the same way as the EER analysis (\S \ref{sec:ana:spec:fit}), 
the escape photons from the summed point sources were taken into account
by adding their best-fit models, after multiplying with the third-order 
polynomial function. Unlike the EER case as a whole, 
this effect is significant ($\sim$20-30\% of the diffuse emission in 
0.5--7 keV) in a few regions (AXJ, C3, and CB) hosting very bright X-ray 
sources ($\simgt$ 1000 netcounts).

\subsubsection{Soft regions}
\label{ngc6334:chandra:diff:pos:fitting:1t}

\begin{figure}[htbp]

\centerline{
\includegraphics[width=5cm,clip,angle=-90]
%%{fig/axjnew2_excl_ps.eps}
{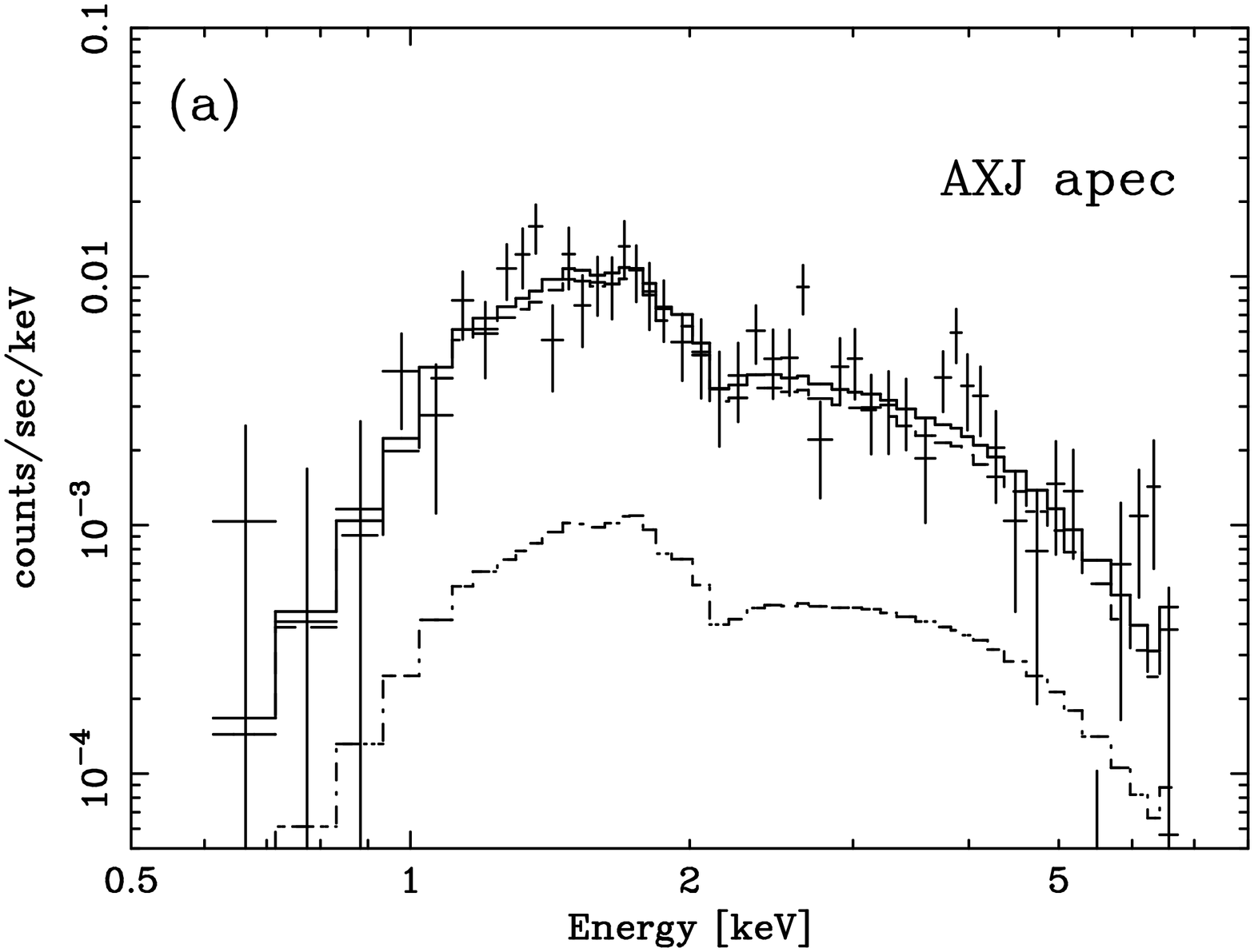}
\hspace*{0.5cm}
\includegraphics[width=5cm,clip,angle=-90]
%%{fig/c2new3_excl_ps.eps}
{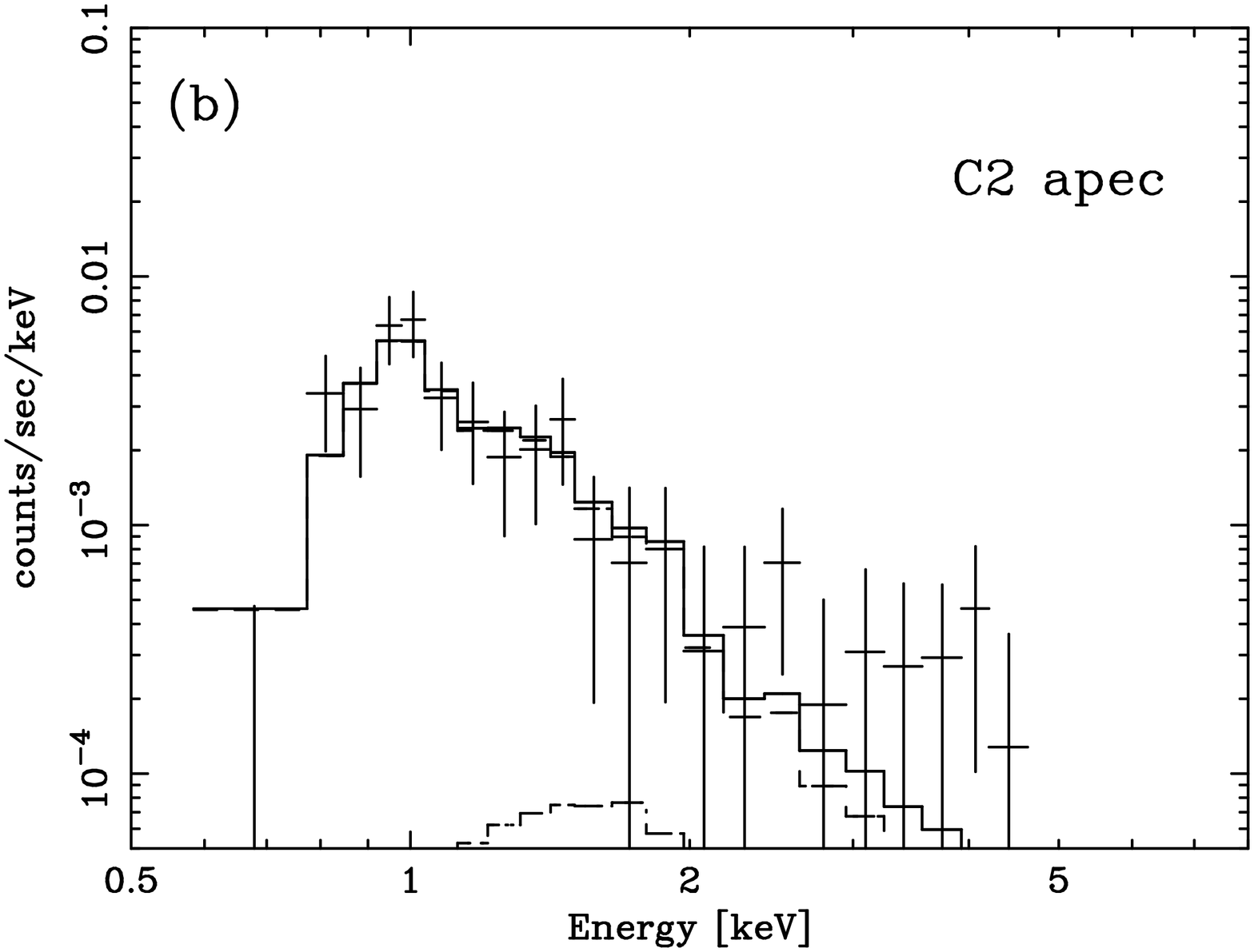}
}
\bigskip
                                                                                                       
\centerline{
\includegraphics[width=5cm,clip,angle=-90]
%%{fig/c3new_r90_excl_ps.eps}
{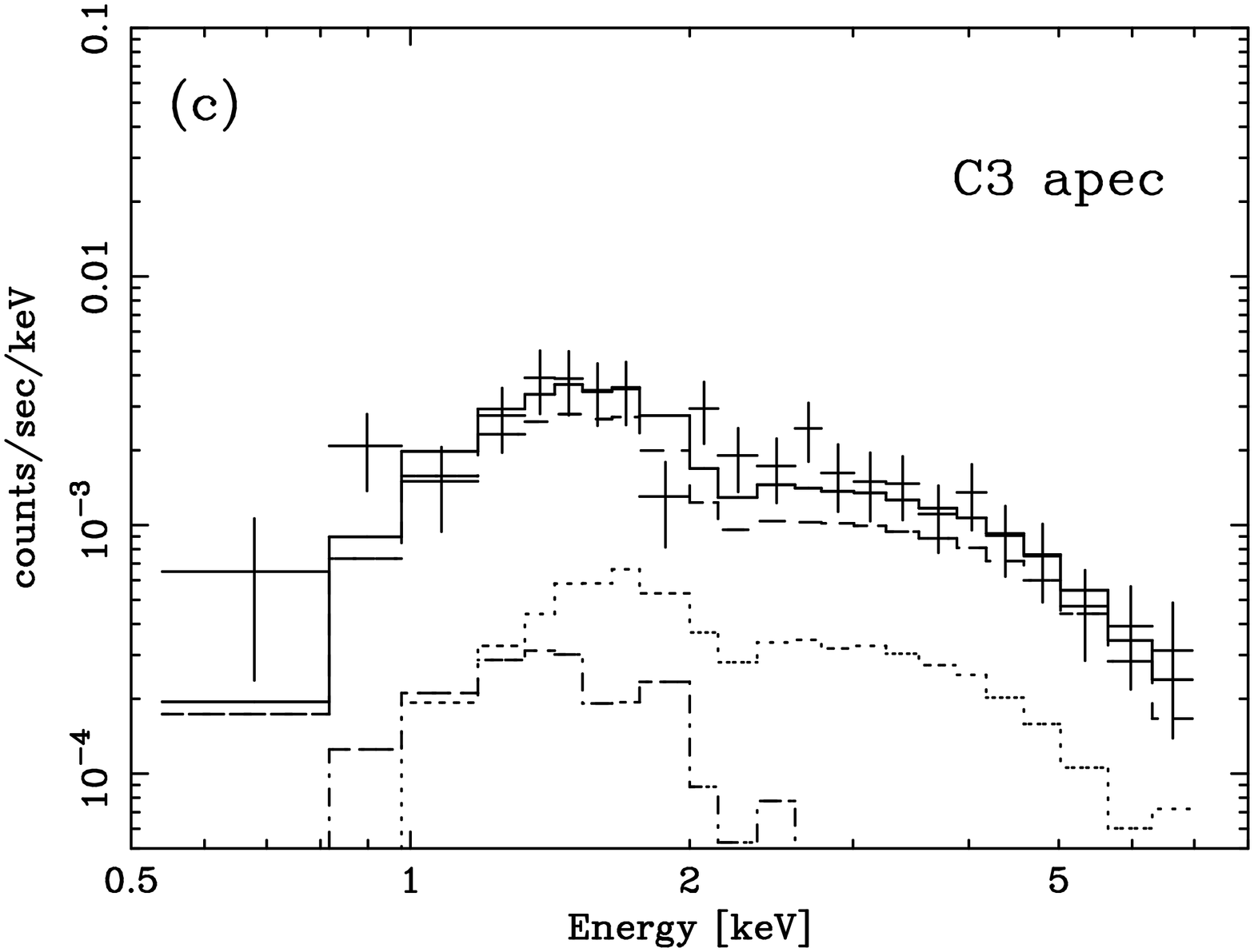}
\hspace*{0.5cm}
\includegraphics[width=5cm,clip,angle=-90]
%%{fig/c4enew_excl_ps.eps}
{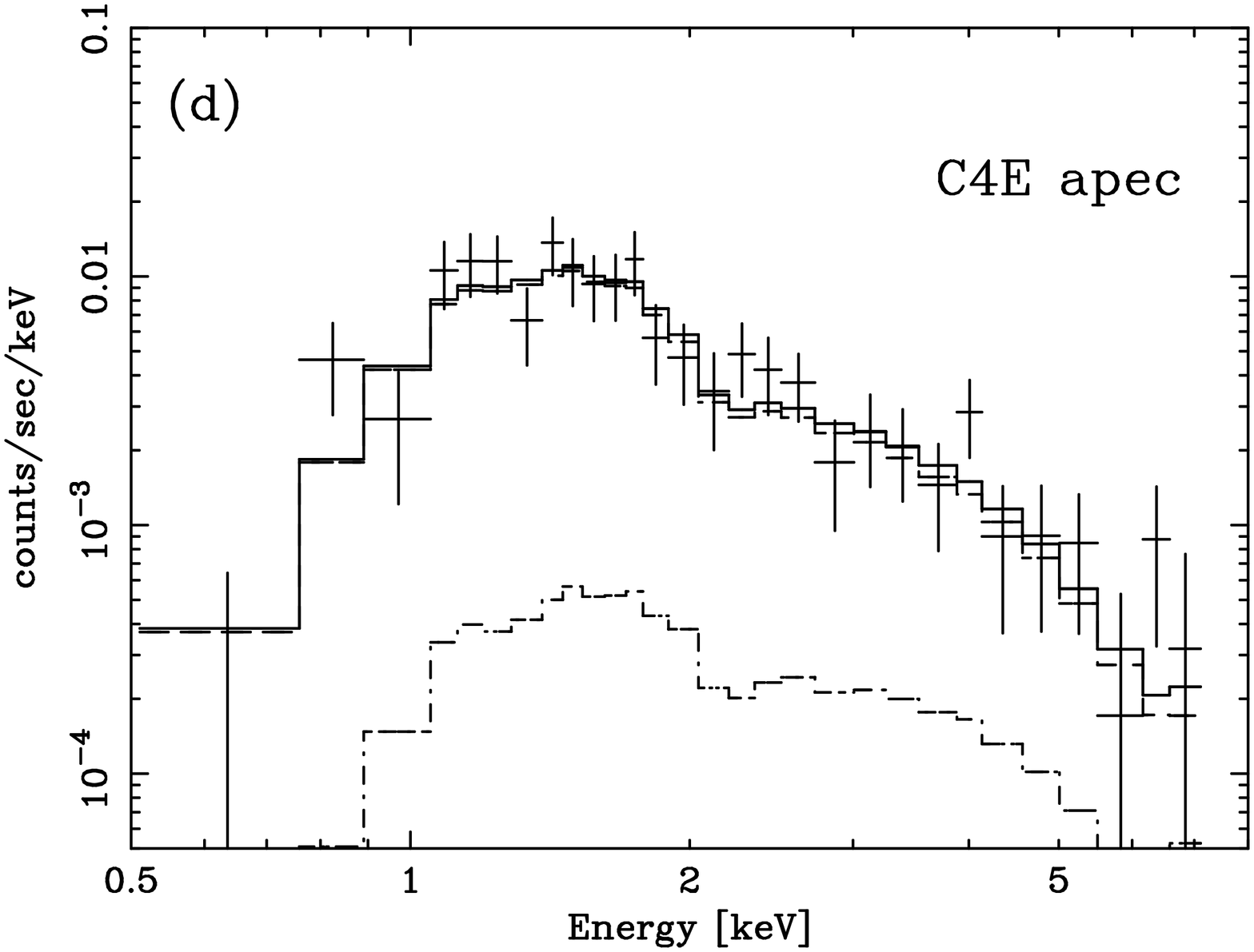}
}
\bigskip
                                                                                                       
\centerline{
\includegraphics[width=5cm,clip,angle=-90]
%%{fig/c5nnew_excl_ps.eps}
{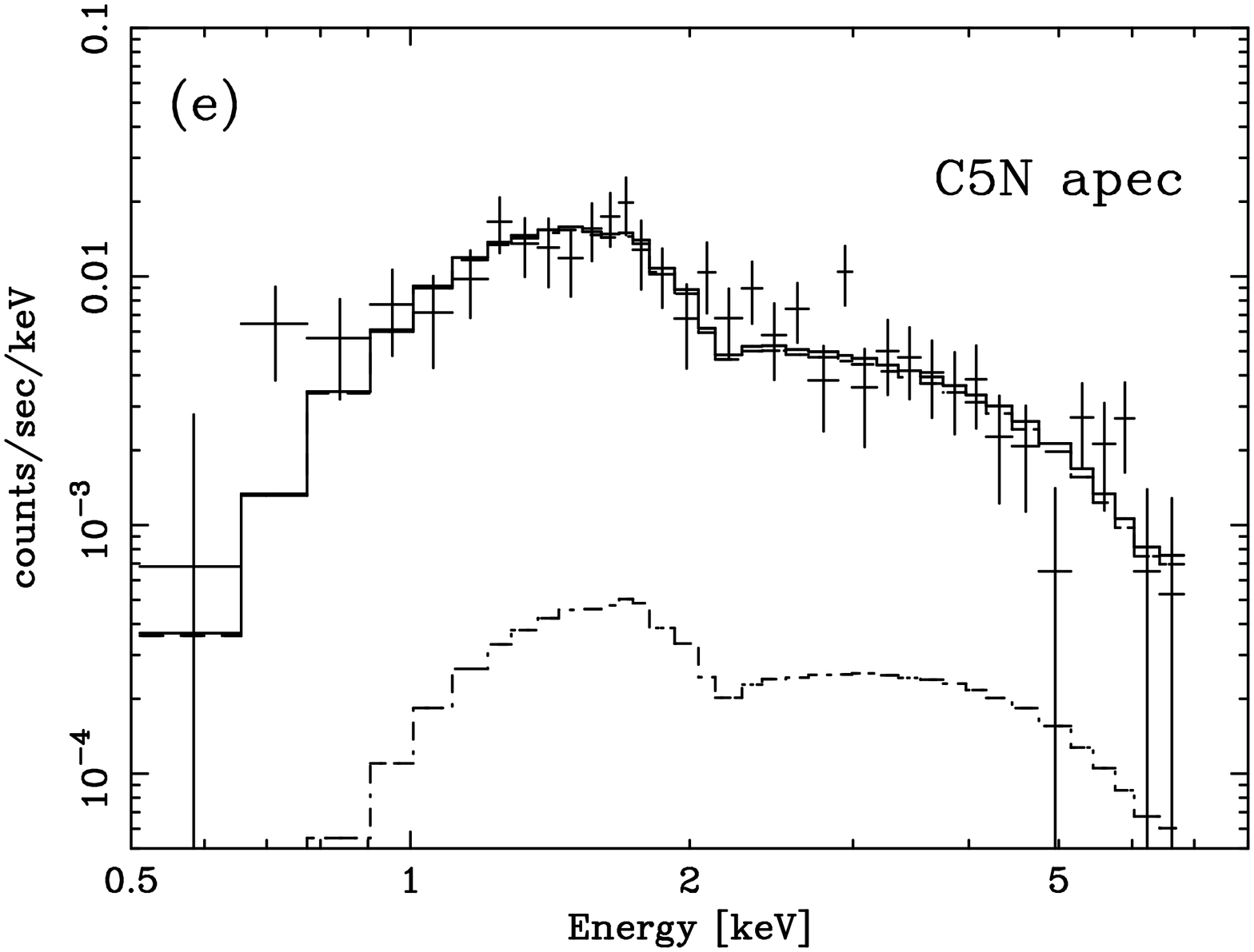}
}
                                                                                                       
\bigskip
                                                                                                       
\caption{The diffuse emission spectra in the 5 soft regions,
         fitted with a model consisting of a thermal plasma emission (dashed),
         and the escape photons from the excluded point sources (dash-dotted).
         See table \ref{tbl:ngc6334:chandra:diff:spec:pos:param1}
         for the obtained parameters.
         }
\label{fig:ngc6334:chandra:diff:spec:pos:fit1}
\end{figure}

We fitted the spectra of the 5 soft regions with a single temperature plasma
model. The temperature and the column density were left free to vary, while
the heavy element abundances were fixed at 0.3 solar.
Results are shown in figure \ref{fig:ngc6334:chandra:diff:spec:pos:fit1},
and the derived parameters are listed in 
table \ref{tbl:ngc6334:chandra:diff:spec:pos:param1}.
Thus, the fits have been acceptable in all the 5 cases.

As expected from the color-color diagram, 
all the 5 spectra show relatively low 
absorption column densities, in the range 
of $0.4\sim1.1\times10^{22}$ cm$^{-2}$. The measured absorption 
agrees, within respective errors, with that of the summed point sources
in the same region. The estimated temperatures range from 1 (C2) to 9 keV 
(C5N), except that in C3 which is not well constrained. 
The region C3 needs some caution, because its surface brightness is rather
low (figure \ref{fig:ngc6334:chandra:dmfilth2}a), its diffuse-emission 
temperature is not well determined 
(table \ref{tbl:ngc6334:chandra:diff:spec:pos:param1}), 
and its point-source spectrum requires an additional softer continuum
(figure \ref{fig:ngc6334:chandra:diff:spec:srcsum:summary}).
Therefore, the absorption column density in this region may still
take multiple values.

\subsubsection{Hard regions}
\label{ngc6334:chandra:diff:pos:fitting:po}

In the same way, we fitted all the 7 hard-region spectra with a single 
temperature model. Because photon statistics in the soft X-ray band are 
rather limited and the $HR_1$ in the color-color diagram 
(figure \ref{fig:ngc6334:chandra:diff:pos:hr1hr2-2})
are similar between the diffuse emission and the summed point sources, 
we fixed the absorption column densities at those of the summed 
point sources ($3.5\sim8.8\times10^{22}$ cm$^{-2}$, see the first line 
in table \ref{tbl:ngc6334:chandra:diff:spec:pos:param2}), 
except in the C4B region in which we left  $N_{\rm H}$
free to vary. The abundance was fixed at 0.3 solar, except in C1 and C1S 
for which it was left free to reproduce a sign of Fe-K line emission. 

\begin{figure}[htbp]

\centerline{
\includegraphics[width=5cm,clip,angle=-90]
%%{fig/c1new_excl_ps.eps}
{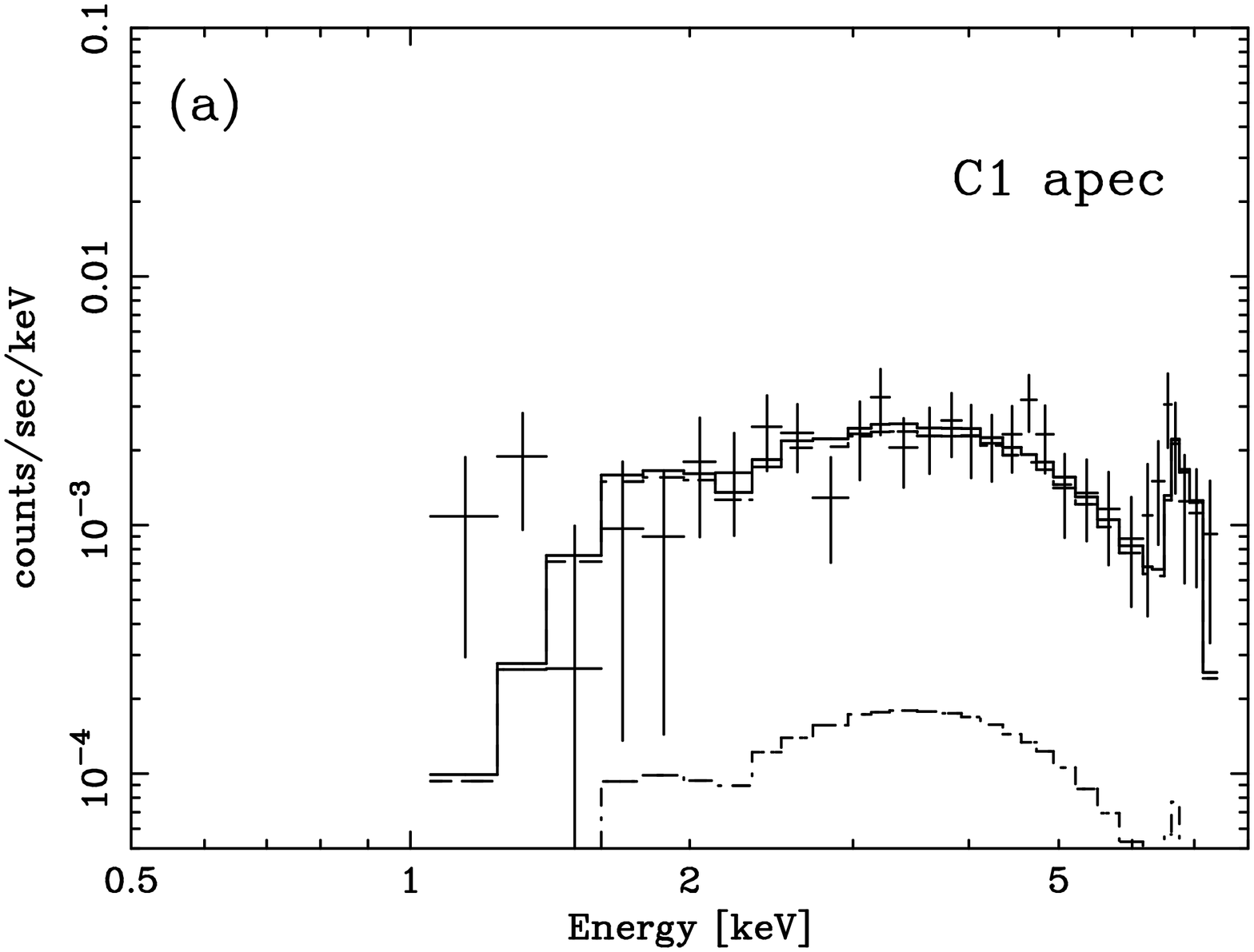}
\hspace*{0.5cm}
\includegraphics[width=5cm,clip,angle=-90]
%%{fig/c1n_excl_ps.eps}
{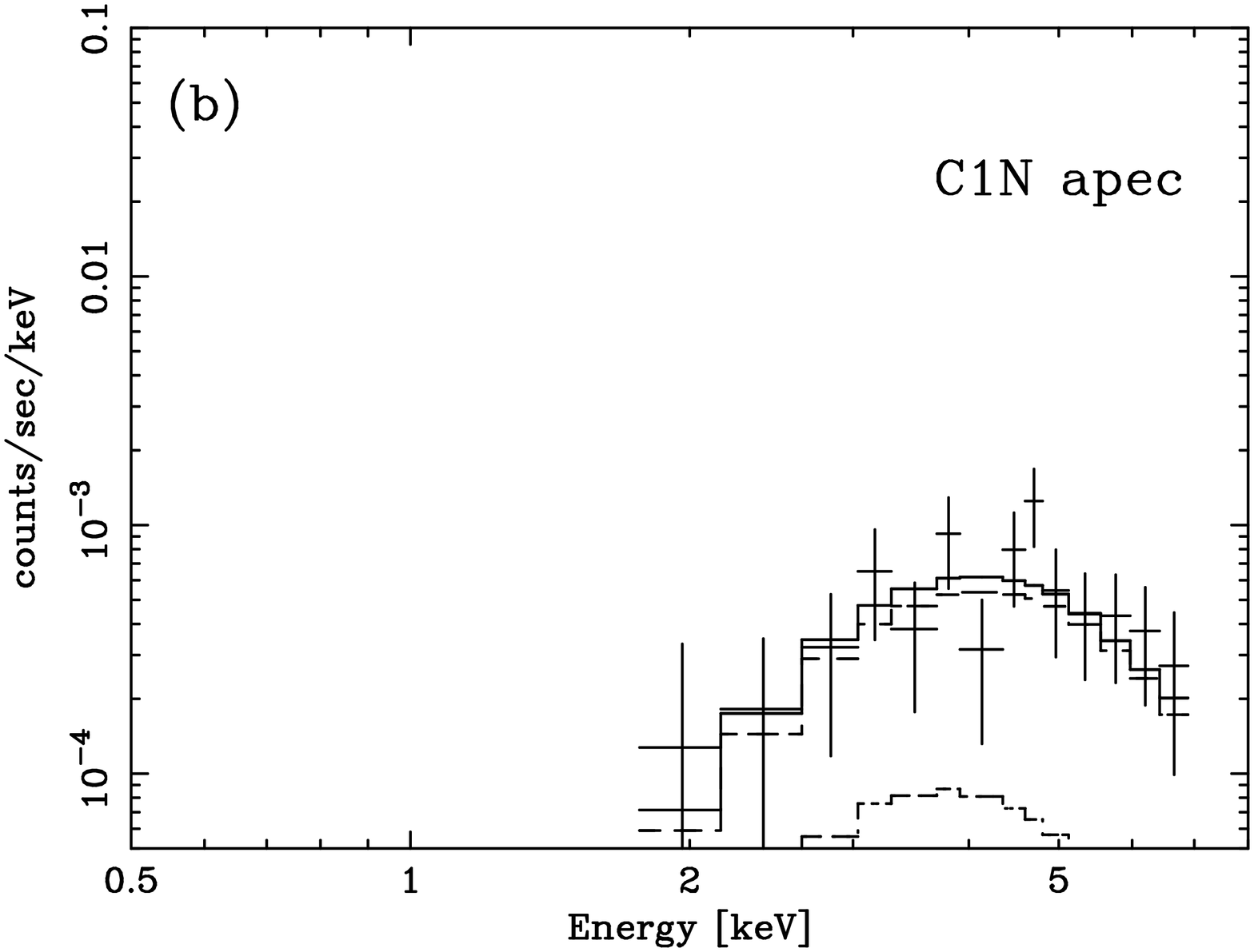}
}
\bigskip
\centerline{
\includegraphics[width=5cm,clip,angle=-90]
%%{fig/c1s_excl_ps.eps}
{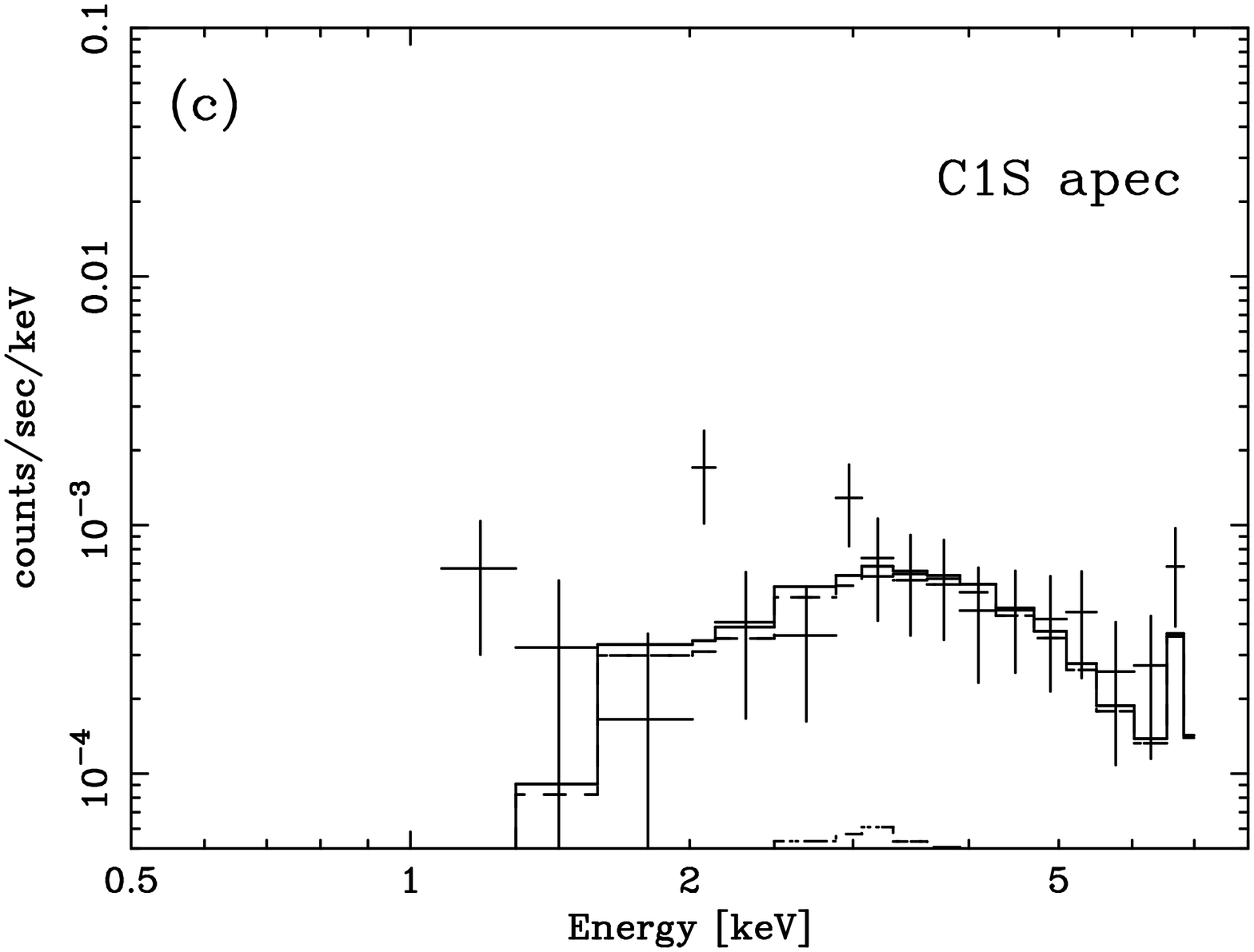}
\hspace*{0.5cm}
\includegraphics[width=5cm,clip,angle=-90]
%%{fig/c2wnew3_excl_ps.eps}
{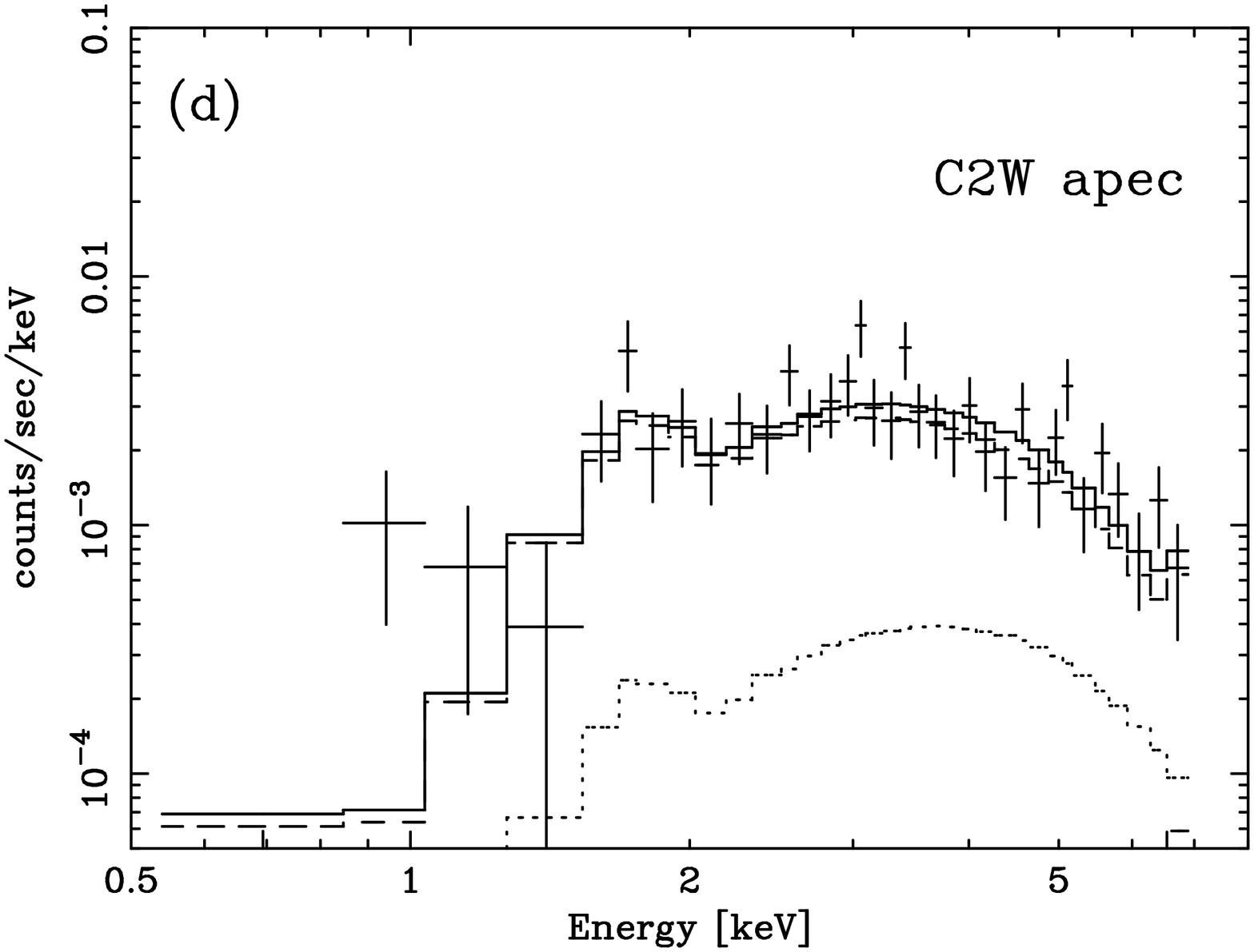}
}
\bigskip

\centerline{
\includegraphics[width=5cm,clip,angle=-90]
%%{fig/c4_excl_ps.eps}
{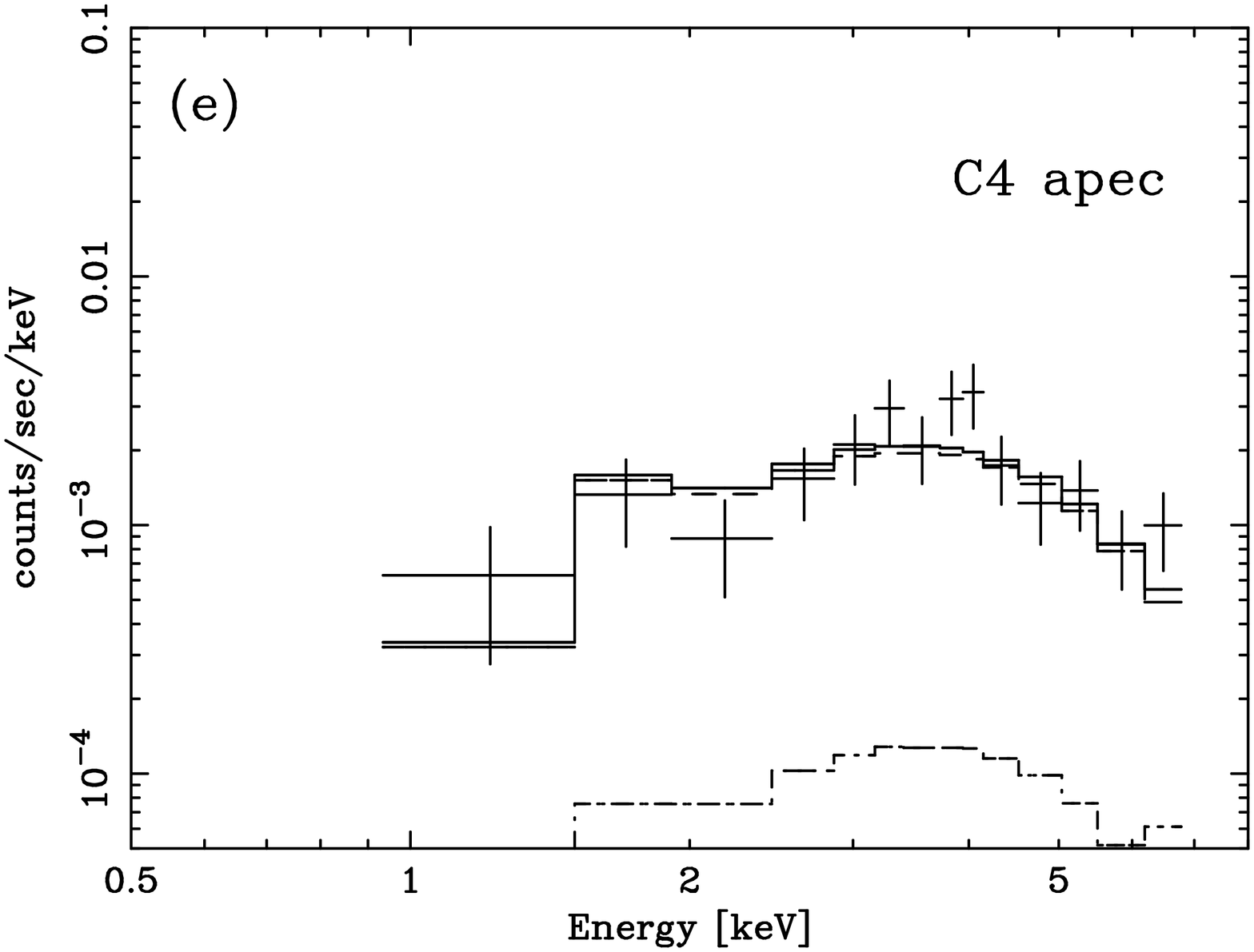}
\hspace*{0.5cm}
\includegraphics[width=5cm,clip,angle=-90]
%%{fig/cb_excl_ps.eps}
{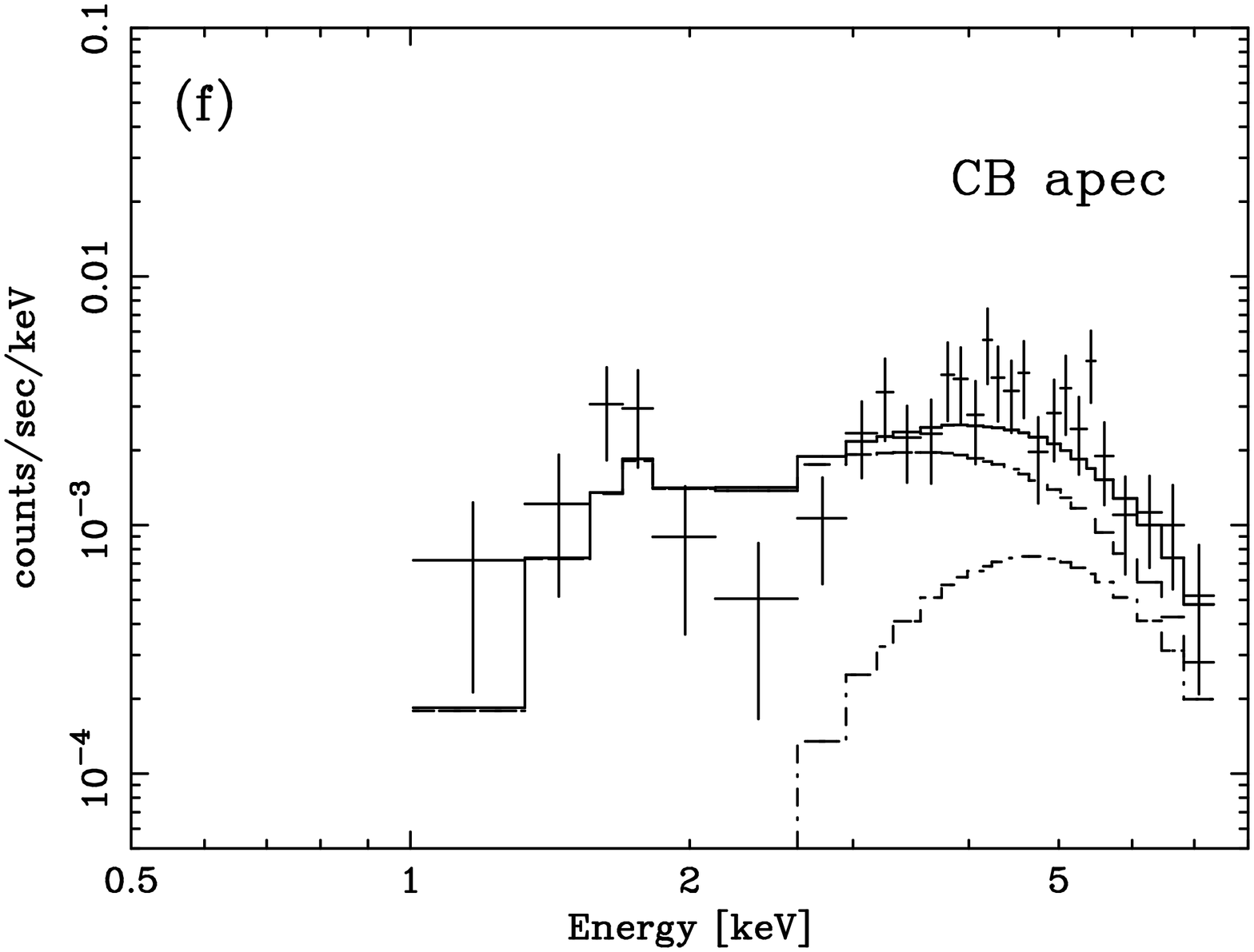}
}
\bigskip

\centerline{
\includegraphics[width=5cm,clip,angle=-90]
%%{fig/c4b_excl_ps.eps}
{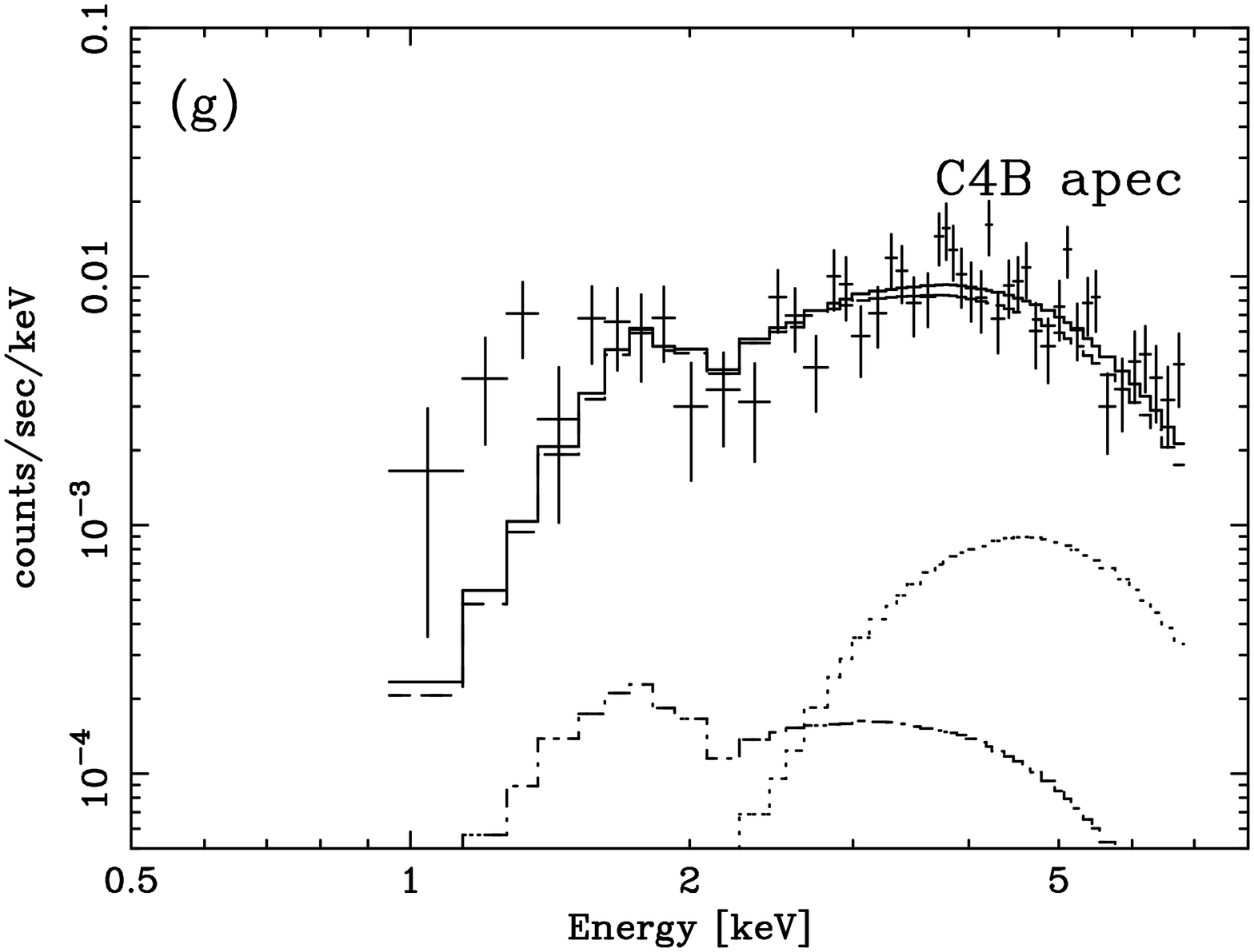}
}
\bigskip
\caption{The same as figure \ref{fig:ngc6334:chandra:diff:spec:pos:fit1} but
         for the 7 hard regions. A narrow Gaussian model is included in C1 and C1S.
         See table \ref{tbl:ngc6334:chandra:diff:spec:pos:param2} for the obtained parameters.
         }
\label{fig:ngc6334:chandra:diff:spec:pos:fit2}
\end{figure}

The results of this analysis are shown in figure 
\ref{fig:ngc6334:chandra:diff:spec:pos:fit2} and table 
\ref{tbl:ngc6334:chandra:diff:spec:pos:param2}. 
All the fits have been acceptable, although that of C4B fit is 
marginal ($\chi^2/\nu\sim1.4$) because of the data excess in 1--2 keV. 
As already suggested by the color-color
diagram, the diffuse continua in the hard regions are generally
flatter than those in the soft regions. Nevertheless, the C1, 
C1N, C1S and C2W spectra could still be regarded as dominated by thin
thermal emission, because the obtained temperatures (5$\sim$10 keV) are 
not unusual among cosmic hot plasmas, and the C1 spectra (and possibly
C1S spectra too) shows the emission feature attributable to Fe-K lines.

Among the four spectra (C1, C1N, C1S, and C2W) with reasonable temperatures,
those of C1N, C1S, and C2W show reasonable abundances (solar or sub-solar).
However, the C1 spectrum requires a very high ($>$1.9 solar) abundance, because
of the strong Fe-K line. If the spectrum is fitted by a phenomenological model
consisting of a power-law continuum and a Gaussian, 
the line center energy is obtained as 6.7$_{-0.2}^{+1.0}$ keV, with the intrinsic
line width of 180 ($<830$) eV and an equivalent width was as huge as 1.5 keV. 
The center energy implies a highly ionized Fe-K line. Then, if we naturally 
interpret the C1N spectrum as thermal emission, the large equivalent width may 
be due to a high local abundance. On the other hand, as suggested in 
the Galactic ridge X-ray Emission \citep{Masai2002}, the hard 
continuum of C1N and also the Fe K-line may partially be quasi-thermal,
i.e., arising from Coulomb collisions of accelerated (non-thermal) electrons 
with thermal ions. This quasi-thermal component can create an apparently 
thermal spectrum of several keV and increase the equivalent width of 
the ionized Fe-K line. Hence, we cannot reject this non-thermal 
possibility for the C1N spectrum.

\begin{figure}[htbp]
\centerline{
\includegraphics[width=5cm,clip,angle=-90]
%%{fig/c4_excl_ps2.eps}
{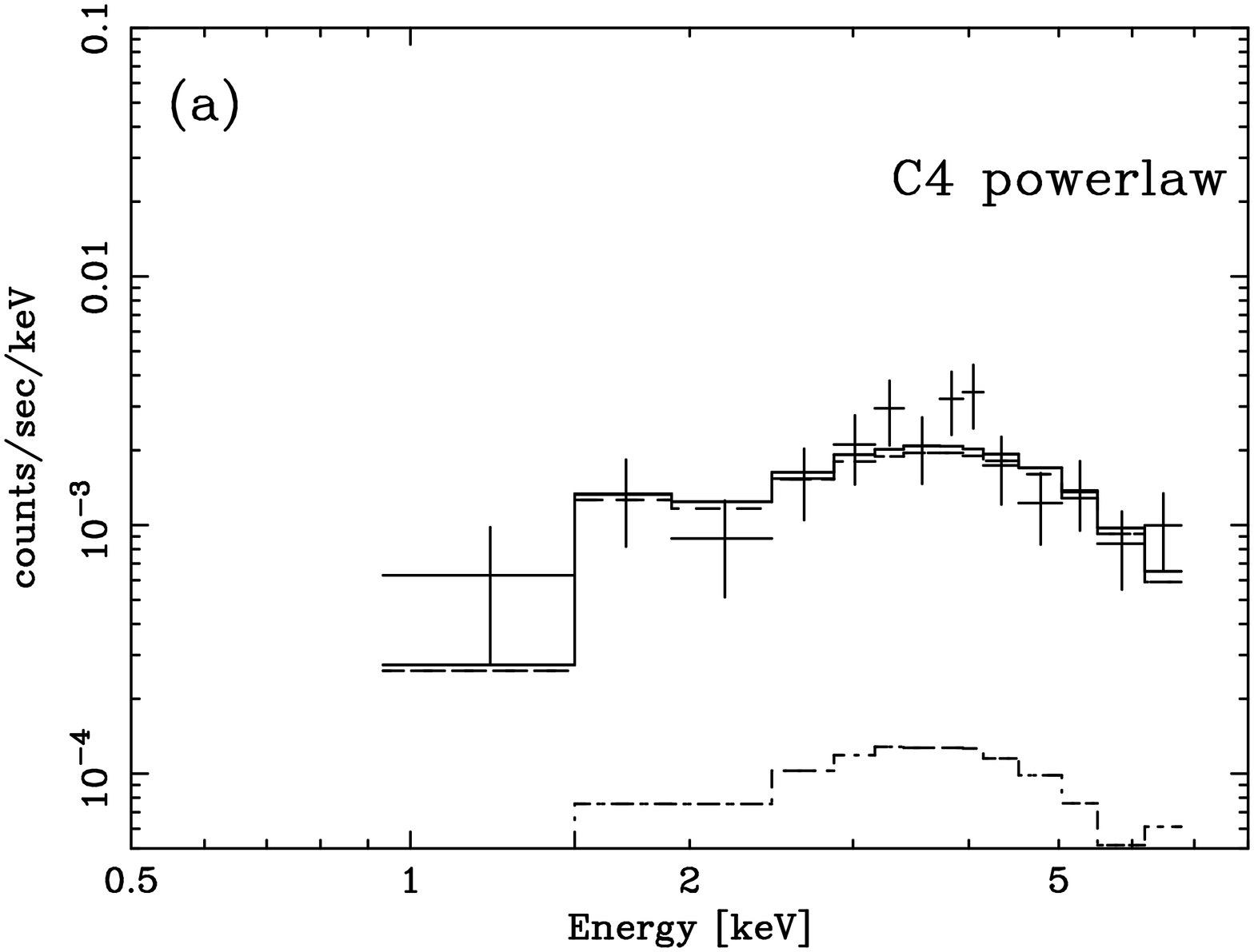}
\hspace*{0.5cm}
\includegraphics[width=5cm,clip,angle=-90]
%%{fig/cb_excl_ps2.eps}
{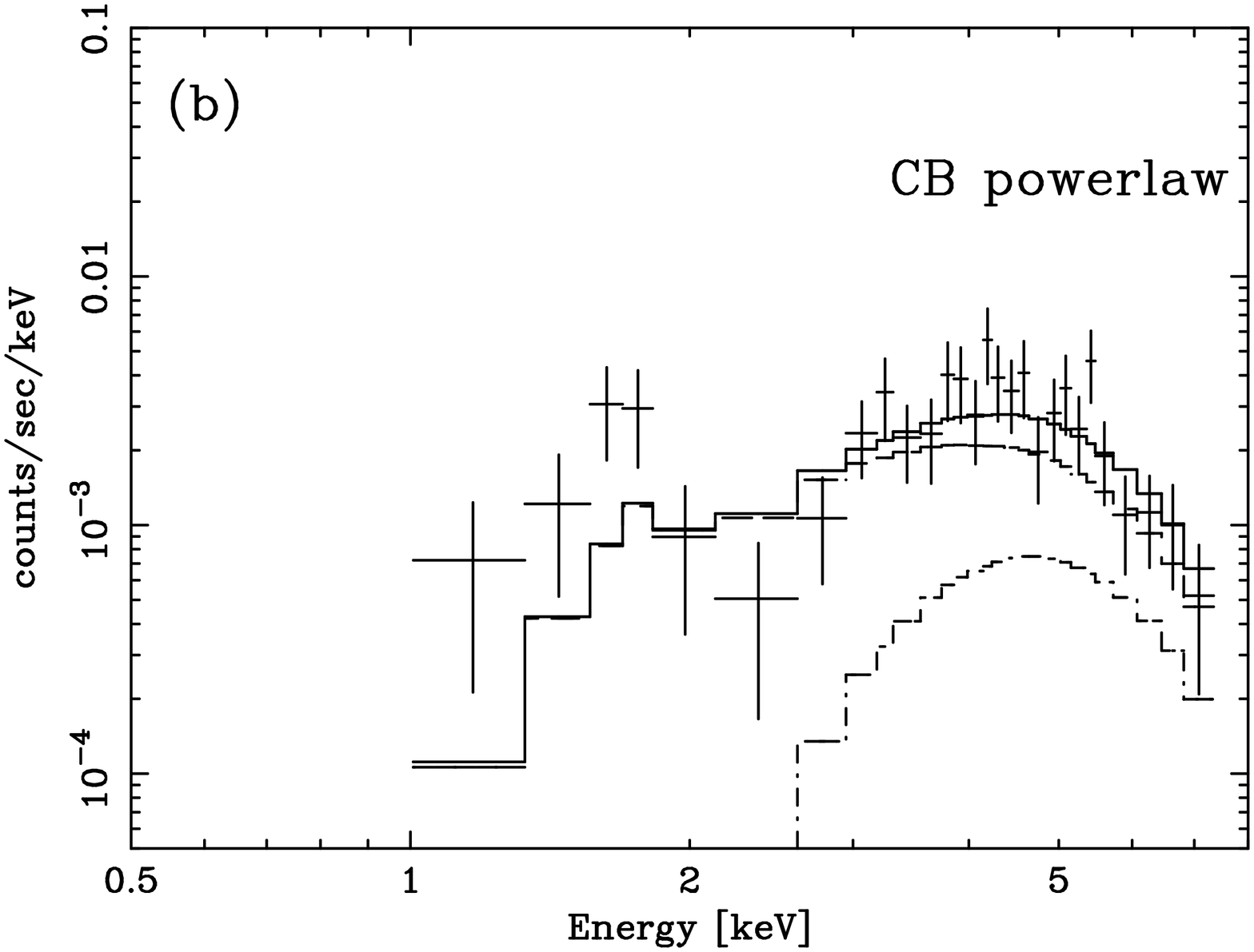}
}\bigskip

\centerline{
\includegraphics[width=5cm,clip,angle=-90]
%%{fig/c4b_excl_ps2.eps}
{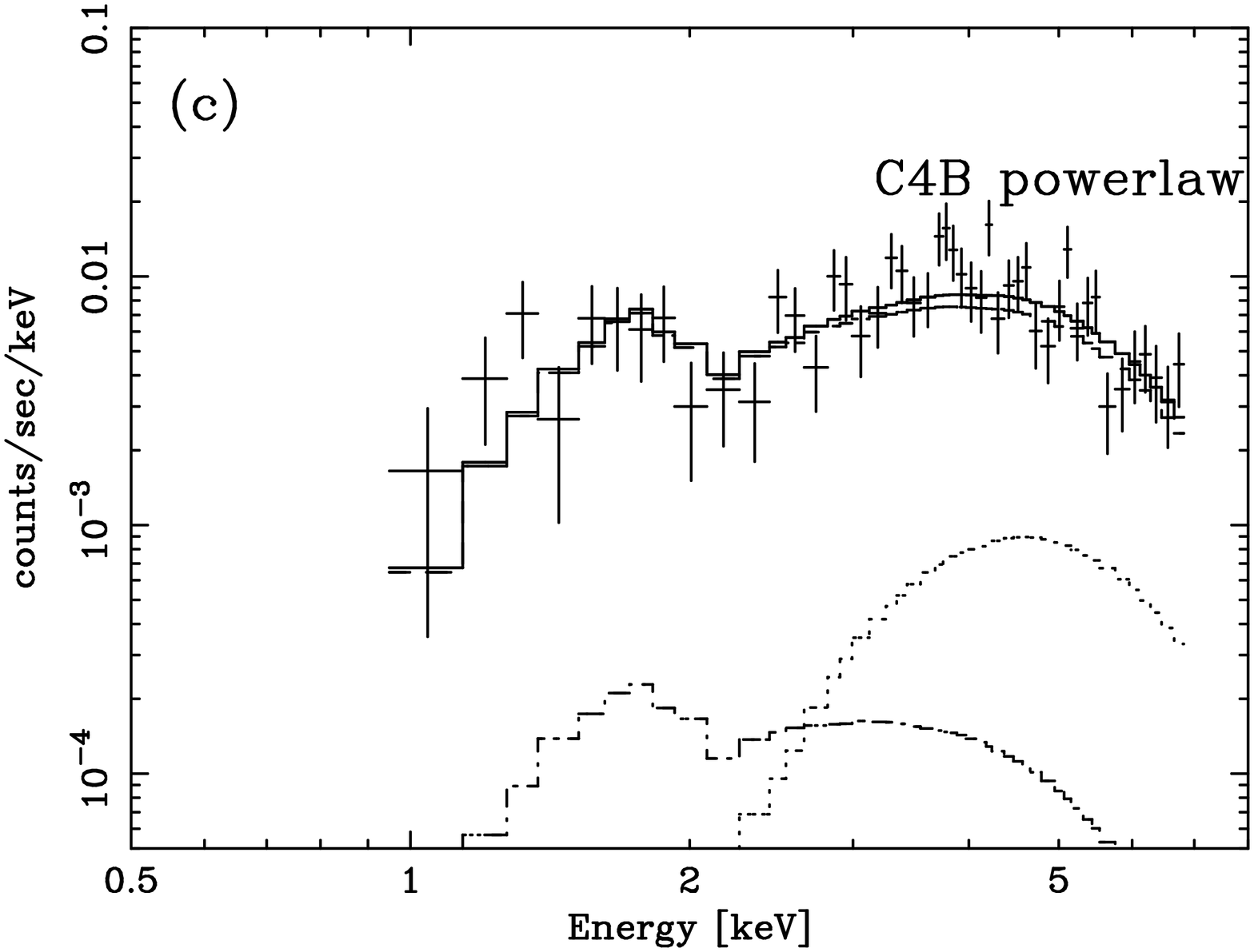}
\hspace*{0.5cm}
\includegraphics[width=5cm,clip,angle=-90]
%%{fig/c4b_excl_ps3.eps}
{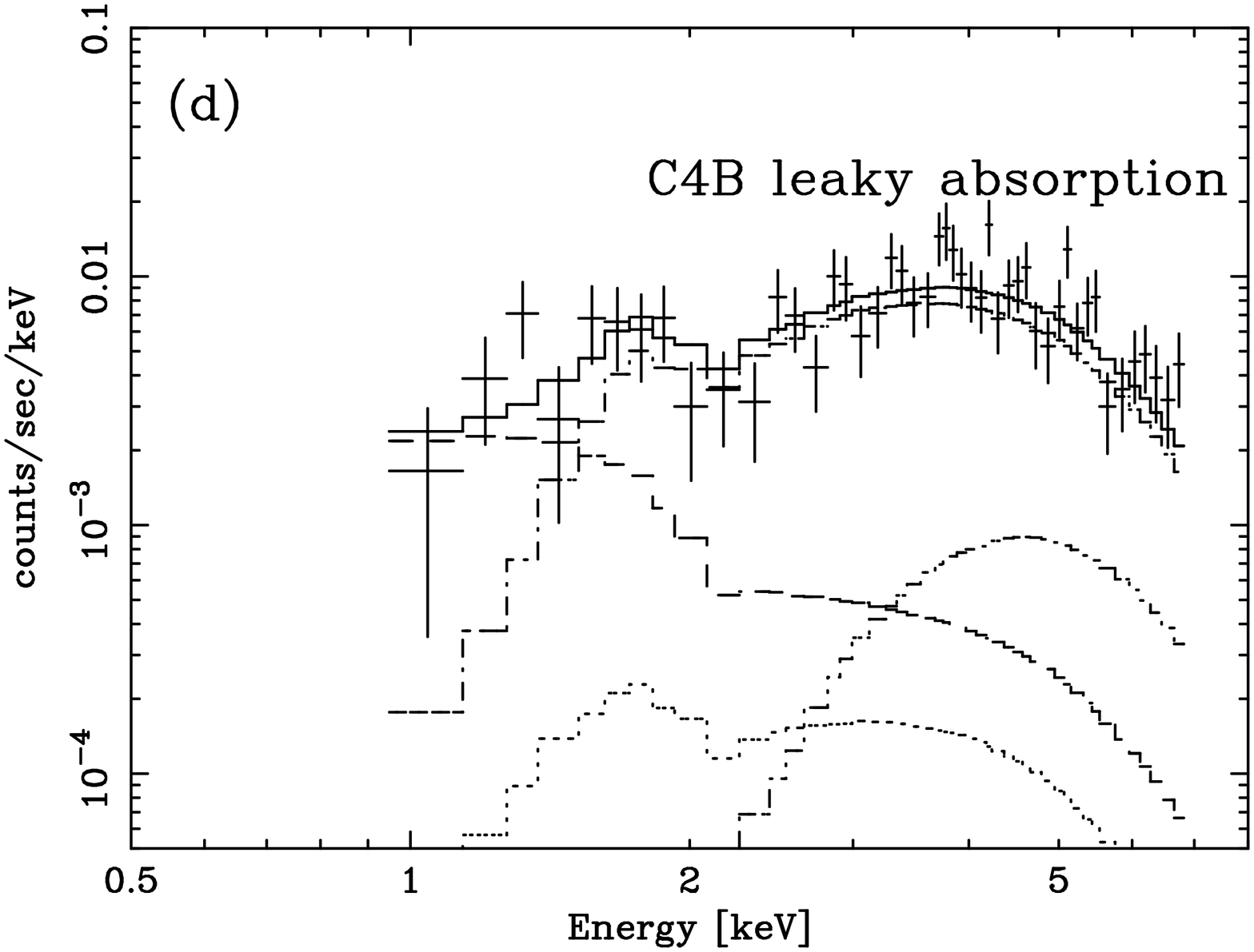}
}
\bigskip
\caption{The same as figure \ref{fig:ngc6334:chandra:diff:spec:pos:fit2} but
         for the C4, CB and C4B spectra fitted with a power-law model (panels a
         through c), and the same C4B spectra fitted 
         with a leaky-absorber thermal model (panel d).
         See table \ref{tbl:ngc6334:chandra:diff:spec:pos:param3} 
         and \ref{tbl:ngc6334:chandra:diff:spec:pos:param4} for the obtained
         parameters.
         }
\label{fig:ngc6334:chandra:diff:spec:pos:fit3}
\end{figure}

The remaining three spectra (C4, CB, and C4B) lack emission
lines and exhibit very flat continua, requiring $kT>10$ keV
if adopting thermal interpretation. Such flat continua could
be better interpreted as non-thermal emission rather than thermal
signals. Accordingly, we refitted them by a power-law model, and
actually obtained an comparable or even better fits as shown  
in figure \ref{fig:ngc6334:chandra:diff:spec:pos:fit3}.
Furthermore the data excess in the 1--2 keV range, which was
observed in the thermal-fit to the C4B spectrum, has disappeared 
because of a decrease in $N_{\rm H}$. 
The obtained power-law indices are extremely small ($0\sim1$) 
with 90\% confidence upperlimits of 1$\sim$1.4. 

Although the ``soft excess'' of the C4B fit in figure
\ref{fig:ngc6334:chandra:diff:spec:pos:fit2} has been
removed by the power-law modeling, it could alternatively
be a result of ``leaky absorber'' condition;
a single thermal emission component reaches us via two (or more)
paths with different absorptions. This is particularly 
likely to be the case with diffuse emission. We hence fitted
the C4B spectrum by a sum of two thermal components with
independent absorptions, but with their temperatures tied
together.
The higher absorption was fixed at the 90\% upper-limit
value of the summed point sources ($4.0\times10^{22}$ cm$^{-2}$).
Then, the soft excess has been explained away by the less absorbed
($<$1$\times10^{22}$ cm$^{-2}$) component. However, the common
temperature has still remained unrealistically high ($>$30 keV).
When the higher absorption is left free to vary, the temperature 
decreased to 4 keV but the higher absorption increases to 
1.1$\times10^{23}$ cm$^{-2}$, which is 3 times higher than the upper-limit
absorption obtained from the summed point sources in the same region.
Thus, the leaky-absorber assumption does not relax the extreme
requirements (a high temperature and a high absorption) for
the thermal interpretation.

Finally, we fitted the C4B spectrum by a sum of a power-law and
a thermal model, modified by separate absorptions. The results
are basically the same as the leaky-absorber model. The power-law
component still showed a small photon index ($1.2$) and a strong
absorption ($5\times10^{22}$ cm$^{-2}$)
to reproduce the hard continuum, while the plasma model with
a low temperature of $kT\sim0.3$ keV and a lower absorption
($2\times10^{22}$ cm$^{-2}$) to explain the soft excess.
Hence, the most favored interpretation of the C4B spectrum is 
still the single power-law model.

\section{Discussion}
\label{sec:discussion}

\subsection{Emission Mechanism and Energy Supply}
\label{discuss:em}

We detected extended hard X-ray emission from this representative MSFR, 
and found that it is likely to be dominated by truly diffuse emission,
rather than formed mainly by unresolved faint point sources.
Taking it for granted that the emission is of diffuse nature, we
showed that it may well be a mixture of 
thermal and non-thermal components. Below we examine the two possible
emission mechanism, and estimate the necessary energy supply.

\subsubsection{Non-thermal interpretation}
\label{discuss:em:nt}

We have found flat continua in some part of the diffuse 
emission of NGC 6334 (C4, CB, and C4B regions). The best
example is the C4B spectrum, which 
has a photon index of $\Gamma=0.39_{-0.63}^{+0.66}$ 
with a 0.5--8 keV luminosity of $4\times10^{32}$ \ergs.
As mentioned in \S \ref{ngc6334:chandra:diff:pos:fitting:po},
such flat spectra are more reasonably interpreted
as non-thermal emission than thermal signals. Taking
this for granted, there can be three candidate emission mechanisms;
bremsstrahlung from non-thermal ($>$10 keV) electrons, 
inverse Compton scattered emission from hundreds-MeV 
electrons, and synchrotron emission from multi-TeV electrons. 

Among the three candidates, we can easily rule out the 
synchrotron emission based on the observed flat X-ray spectra. 
In fact, synchrotron emission spectra, observed in various
frequencies from various objects 
(e.g., supernova remnants or SNRs, \citealt{Koyama1995};
Blazers, \citealt{Kubo1998}),
all exhibit photon indices steeper than 1.5.
This is consistent with the generally accepted view that the highly
energetic electrons responsible for the synchrotron emission are produced
by the standard diffusive shock acceleration mechanism \citep{Longair1994},
which predict electron spectra with indices $\simgt2$ and hence
the emergent photon indies of $\simgt1.5$.
Furthermore, the electrons must have energies exceeding $\sim10^{12}$
eV, in order to emit synchrotron X-rays under a typical magnetic
field strength of $\sim100\mu$ G of the cloud cores \citep{Sarma2000}.  
Such extreme electron energies are difficult to explain.

If assuming the inverse-Compton process, the requirement on the
electron energy can be much relaxed, because electrons with energies
of several hundred MeV (with a Lorenz factor of $\sim10^{2-3}$) can 
produce X-ray photons by scattering off strong IR lights
that dominates in the environment considered here. However, the 
observed flat spectrum makes this interpretation also unlikely
for the same reason, because the inverse-Compton emission 
from a population of energetic electrons has the same spectral slope
as their synchrotron spectrum.

The remaining possibility is thus the bremsstrahlung emission
by mildly energetic electrons. Unlike the former two mechanisms,
the electron energy only have to exceed $\sim10$ keV which is the highest
photon energies observed.
In an environment with a high matter density reaching $10^{3-4}$ cm$^{-3}$
like in these two cases, the bremsstrahlung loss overwhelms the synchrotron
and inverse-Compton losses up to electron energies of $\sim10$ MeV, and
the spectrum of energetic electrons will become flatter
than the initially injected spectrum because
lower-energy electrons lose energy more
quickly than the more energetic ones via the Coulomb loss.
This makes a contrast to the synchrotron and inverse-Compton mechanisms,
in which the energy loss always steepen the electron spectrum.
Such electrons will emit bremsstrahlung with a photon index $\Gamma=1$,
as is expected for a mono-energetic case. Actually,
\citet{Uchiyama2002} successfully explained a flat
($\Gamma\sim1$) photon spectrum observed from hard X-ray clumps
toward the SNR $\gamma$ Cygni,
in terms of this process invoking tens MeV electrons.
In our case, the observed flat spectra may be explained
in the same manner, and hence the bremsstrahlung emission is the
most favorable at least from the viewpoint of spectral shapes.

Although the bremsstrahlung interpretation is feasible from several
important aspects, one issue remains; in this energy range, 
the Coulomb loss overwhelms the bremsstrahlung 
emission by a factor of $\sim10^{3-5}$. 
Therefore,
the bremsstrahlung emission is possible, only if a kinetic
luminosity of more than $10^{36}$ \ergs is supplied to the 
hard X-ray clump at the C4B region.  
This is examined in \S \ref{discuss:stellarwind:energy}.

\subsubsection{Thermal interpretation}
\label{discuss:em:th}

Since some parts (particularly the soft regions) of the EER
are thought to be emitting optically-thin thermal X-rays, we may
also examine whether the thermal interpretation is physically
feasible. The bolometric luminosity $L_{\rm X}$ of a thin-thermal
plasma of a temperature $T$ is expressed, in terms of 
the electron density $n_{\rm e}$ and the emitting volume $V$ as,
\begin{equation}
\displaystyle
\begin{array}{ccccc}
\mbox{$L_{\rm X}$}         & =  & \mbox{$\Lambda (T) EM $} 
                           & =  & \mbox{$\Lambda (T) {n_{\rm e}}^2 V \eta$},\\
\end{array}
\label{eqn:discuss:EM}
\end{equation}
\noindent
where $\Lambda$, $EM$, and $\eta$ 
denote the cooling function, 
the volume emission measure, and a filling factor 
of the emitting plasma ($\leq1$), respectively. 

From the observations, we here assume $kT=5$ keV, 
$L_{\rm X}=5\times10^{31}$ \ergs, and $V=\frac{4\pi}{3}r^3$ 
($r=$0.5 pc). Also we approximate 
the cooling function of a low density plasma of a solar 
metallicity as 
%%$3.6\times10^{-23}\cdot(kT)^{-0.6}$ \ergs cm$^{3}$ s$^{-1}$ 
%%for $kT=$0.01--3 keV 
%%and
$1.0\times10^{-23}\cdot(kT)^{0.5}$ \ergs cm$^{3}$ s$^{-1}$ 
which is valid for $kT>$3 keV \citep{Raymond1976,McKee1977}.
Then, equation (\ref{eqn:discuss:EM}) is solved 
for $n_{\rm e}$ as
\begin{equation}
\displaystyle
\begin{array}{ccl}

%%\mbox{$EM$} &=& 
%%  \mbox{2$\times10^{54}$} %%~\beta 
%%  ~{\rm cm}^{-3}~
%%  (\frac{\mbox{$L_{\rm X}$}}{\mbox{$5\times10^{31}$ergs s$^{-1}$}}),\\

\mbox{$n_{\rm e}$} &=& 
\mbox{0.4} %%~\beta^{\frac{1}{2}} 
~\eta^{-\frac{1}{2}} ~{\rm cm}^{-3}~
(\frac{\mbox{$L_{\rm X}$}}{\mbox{$5\times10^{31}$ergs s$^{-1}$}})^{\frac{1}{2}}
(\frac{\mbox{$r$}}{\mbox{$0.5$ pc}})^{-\frac{3}{2}}.\\

\end{array}
\label{eqn:discuss:ne}
\end{equation}

\noindent
The total plasma energy $U$ of a single soft region, 
the plasma pressure $p$, and the radiative 
cooling time scale $t_{\rm cool}$ are then derived as
\begin{equation}
\displaystyle
\begin{array}{lcccl}

\mbox{$U$} & =  & \mbox{$3n_{\rm e} kT V$}& =  &
\mbox{$1\times10^{47}$} 
 ~\eta^{\frac{1}{2}}~{\rm ergs}~
\left(
\frac{\mbox{$L_{\rm X}$}}{\mbox{$5\times10^{31}$ergs s$^{-1}$}}
\right)^{\frac{1}{2}}
\left(
\frac{\mbox{$kT$}}{\mbox{$5$ keV}}
\right)
\left(
\frac{\mbox{$r$}}{\mbox{$0.5$ pc}}
\right)^{\frac{3}{2}},\\

\end{array}
\label{eqn:discuss:U}
\end{equation}
\noindent
\begin{equation}
\displaystyle
\begin{array}{lcccl}

\mbox{$p$} & =  & \mbox{$2n_{\rm e} kT  $} & =  &
\mbox{$4\times10^{7}$} %%~\beta^{\frac{1}{2}} 
 ~\eta^{-\frac{1}{2}}~{\rm K~cm}^{-3}~
\left(\frac{\mbox{$L_{\rm X}$}}{\mbox{$5\times10^{31}$ergs s$^{-1}$}}\right)^{\frac{1}{2}}
\left(\frac{\mbox{$kT$}}{\mbox{$5$ keV}}\right)
\left(\frac{\mbox{$r$}}{\mbox{$0.5$ pc}}\right)^{-\frac{3}{2}},\\

\end{array}
\label{eqn:discuss:p}
\end{equation}
and
\noindent 
\begin{equation}
\displaystyle
\begin{array}{lcccl}

\mbox{$t_{\rm cool}$} & =  & \mbox{$U/L_{\rm X}$} & =  &
\mbox{$9\times10^{7}$} %%~\beta^{-\frac{1}{2}} 
 ~\eta^{\frac{1}{2}}~{\rm yr}~
\left(\frac{\mbox{$L_{\rm X}$}}{\mbox{$5\times10^{31}$ergs s$^{-1}$}}\right)^{-\frac{1}{2}}
\left(\frac{\mbox{$kT$}}{\mbox{$5$ keV}}\right)
\left(\frac{\mbox{$r$}}{\mbox{$0.5$ pc}}\right)^{\frac{3}{2}}.\\

\end{array}
\label{eqn:discuss:tcool}
\end{equation}

The cooling time as estimated above is far longer
than the typical age of the massive star-forming 
regions ($10^{5-6}$ yr), and sound crossing time
($\sim10^{3}$ yr) in a 5 keV plasma across the region
of $\sim0.5$ pc in size. Therefore, the total energy $U$ 
must be accumulated over the MSFR age if the plasma is 
confined, and over the sound crossing time otherwise.
Although the thermal plasma pressure of equation (\ref{eqn:discuss:p})
is higher than that of
the surrounding molecular clouds ($10^{5-6}$ K cm$^{-3}$),
the plasma may be confined by the surrounding dense HII region, 
where the pressure is thought 
to be higher ($10^{7}\sim10^{8}$ K cm$^{-3}$; \citealt{Rodriguez1982}). 
The magnetic pressure of the molecular cloud (of the order of $100\mu$ G; 
\citealt{Sarma2000}) may help the confinement.
Then, the total 
energy of equation (\ref{eqn:discuss:U}) can be understood as 
an average luminosity of $3\times10^{33-34}$ \ergs 
over the typical age of $10^5\sim10^6$ yr. 
If scaling it to the whole diffuse
emission of NGC 6334, the necessary energy
input becomes $1\times10^{35-36}$ \ergs, which is 
comparable to that required by the non-thermal interpretation.
These values are in fact upper limits, and can be
lowered as $\propto\eta^{\frac{1}{2}}$ by assuming
a lower filling factor.

\subsection{Stellar Winds as the Energy Source}
\label{discuss:stellarwind}

\subsubsection{Energetics}
\label{discuss:stellarwind:energy}

Then, what explains the huge luminosity up to $\sim10^{36-37}$ \ergs
required by either 
non-thermal or thermal picture ? Since the diffuse emission 
is clearly localized to the massive star-forming sites, 
it must have a close connection to the formation of massive
stars. Though there are many 
energetic phenomena (molecular outflows, jets, and HII regions), 
the most plausible candidate is the fast stellar winds 
from massive OB stars; the stellar wind may collide with the 
ambient gas (e.g., dense HII regions), producing shocked regions
which may become the source of both thermal and non-thermal X-rays.

This interpretation 
has been proposed to explain the soft ($kT\simlt1$ keV) possible 
diffuse emission in M17 and Rosette Nebula \citep{Townsley2003}. 
Already in the {\it ASCA} era, \cite{Matsuzaki1999a} also suggested 
this possibility to explain the high temperature spectra of the 
region-integrated emission from NGC 6334. Below, we re-consider
this scenario based on our new results, keeping in mind with the long-studied
stellar-wind shock theory \citep{Dyson1972,Castor1975, Weaver1977}. 

Observationally, 7 of the 12 diffuse emission regions (AXJ, C1, C1N, C1S, C3, C4, and C4B) involve 
at least one OB star candidates 
(\citealt{Matsuzaki1999a, Loughran1986, Straw1989a, Persi2000}).
Although the other 5 regions (C2, C2W, C4E, CB and C5N) do not
include any OB stars reported in the optical, IR or radio bands,
this does not necessarily mean a difficulty with the stellar wind
scenario. Actually, the shocked region around each OB star is 
expected to be rather asymmetric, because of the strong pressure
gradient across the dark lane. Then, the emission region may appear
rather offset, or even detached, from the central star. Furthermore,
some of these regions may host embedded OB stars or represent an 
interaction among winds from more than one massive stars.

These massive late O- or early B-type stars in individual cores
are expected to emit thick and fast stellar winds,
each supplying a kinematic luminosity of
\begin{equation}
\displaystyle
\begin{array}{cccccc}
\mbox{$L_{\rm w}$} & =  & \mbox{$\frac{1}{2}$} \mbox{$\dot{M}$} \mbox{$v_{\rm w}$}^2
                   & =  & 1\times10^{35} ~ {\rm ergs} ~{\rm s}^{-1}
                         ~\left( \frac{\mbox{$\dot{M}$}}{\mbox{$10^{-7}$ \msolar yr$^{-1}$}} \right) 
                         ~\left( \frac{\mbox{$v_{\rm w}$}}  {\mbox{$2000$ km s$^{-1}$}} \right)^{2}, \\
\end{array}
\label{eqn:discuss:windenergy}
\end{equation}
\noindent
where $\dot{M}$ and $v_{\rm w}$ denote the mass loss 
rate and wind velocity, respectively.
Thus, modest values of $\dot{M}$ and $v_{\rm w}$, used to
normalize equation (\ref{eqn:discuss:windenergy}), would
be sufficient to give $1\times10^{35}$ \ergs per star.

Considering several massive stars, we can readily explain 
the energy input required by the thermal interpretation.
On the other hand, when we consider the non-thermal emission 
arises from high energy particles accelerated at the shock front, 
similar to those seen in some SNRs such as SN 1006 \citep{Koyama1995} 
and G347.3$-$0.5 \citep{Koyama1997}, the situation will be more difficult.
In this shock acceleration case, 
because of the conversion efficiency of the kinematic energy 
into sub-MeV electrons, the required energy ($1\times10^{36}$ \ergs for 
C4B region) increases by at least an order of magnitude \citep{Bykov2000}. 
Hence, the non-thermal interpretation needs a cluster of OB
stars or a faster or more massive wind or wind-wind collisions. 
The most plausible case will be the OB cluster and/or wind-wind collisions, 
because at least two late O to early B star candidates are detected 
within C4B region \citep{Straw1989a,Persi2000}. 
Deep NIR to FIR observations with, e.g., {\it Spitzer} is 
necessary to the OB star population in this region.

\subsubsection{Shock temperature}
\label{discuss:stellarwind:temperature}

The wind will experience a shock transition at a certain radius from 
the star, where the ram pressure of the wind becomes equal to the external
pressure (e.g., of HII regions). The shell region between the shock and the
contact discontinuity is then filled with shocked hot winds or hot bubble, 
and become the diffuse thermal X-ray source. The maximum temperature 
$kT_{\rm s}$
behind the shock is given by
\begin{equation}
\displaystyle
\begin{array}{cccccc}
\mbox{$kT_{\rm s}$}&=& \mbox{$\frac{3}{16}$} \mbox{$\mu$} \mbox{$m_{\rm H}$}\mbox{$v_{\rm w}$}^2
                   &=& 5 ~ {\rm keV} ~
                       ~\left( \frac{\mbox{$v_{\rm w}$}}  {\mbox{$2000$ km s$^{-1}$}} \right)^2, \\
\end{array}
\label{eqn:discuss-temperature}
\end{equation}
\noindent
where 
$m_{\rm H}$ is the mass of a hydrogen atom, and $\mu=0.62$ is the mean 
molecular weight. The observed temperatures 
of the soft regions in NGC 6334, 1--10 keV, can be explained
wind velocities in the range of 1000-3000 km s$^{-1}$ as usually 
seen in OB stars \citep{Prinja1990}. The actual temperature behind the shock
may be considerably lower than that eq. (\ref{eqn:discuss-temperature}), 
due to thermal conduction from the shocked hot wind to the surrounding cold 
material \citep{Weaver1977}. 
However, in high stellar density environment of star-forming cores, 
the wind-wind shock may be realized and can increase the shock temperature.

\subsubsection{Wind confinement}
\label{discuss:stellarwind:confinement}

We can estimate the size of expanding hot wind bubble $R_{\rm b}$ 
or the distance from the central OB star to the contact discontinuity
by assuming that the wind energy is equal to the displaced energy of 
the cold gas. This can be described as
\begin{equation}
\displaystyle
\begin{array}{cccccc}
\frac{1}{2} \mbox{$\dot{M}$} \mbox{$v_{\rm w}$}^2 t
& =  & 
\left(\frac{4}{3}\pi {R_b}^3 \right) \frac{3}{2} p_{\rm s},\\
\end{array}
\label{eqn:discuss-wind-bubble}
\end{equation}
\noindent
where $t$ is the time since the wind started blowing,
and $p_{\rm s}$ is the thermal pressure 
of the surrounding material \citep{Chevalier1999}. 
Assuming an HII region with a temperature of 10000 K and
a density of 10$^{3}$ cm$^{-3}$ as the surrounding material, 
and the age of a young massive star as 0.1 Myr, 
we obtain
\begin{equation}
\displaystyle
\begin{array}{cccccc}
R_b & = & 
    1  ~{\rm pc} 
       ~\left( \frac{\mbox{$\dot{M}$}}{\mbox{$10^{-7}$ \msolar yr$^{-1}$}} \right)^{\frac{1}{3}} 
       ~\left( \frac{\mbox{$v_{\rm w}$}}{\mbox{$2000$ km s$^{-1}$}} \right)^{\frac{2}{3}}  
       ~\left( \frac{\mbox{$p_{\rm s}$}}{\mbox{$10^7$ K cm$^{-3}$}} \right)^{-\frac{1}{3}}  
       ~\left( \frac{\mbox{$t$}}  {\mbox{$10^5$ yr}} \right)^{\frac{1}{3}}. 
\\
\end{array}
\label{eqn:discuss-wind-bubble-rb}
\end{equation}
\noindent
This estimation is in a good agreement with the observed size of 
the 12 diffuse emission regions. The whole diffuse emission region of 
NGC 6334 ($\sim5\times9$ pc$^2$) can be explained by a superposition 
(7 or more) of this bubble.

\begin{figure}[htbp]

\centerline{
\includegraphics[width=7cm,clip,angle=-90]
%%{fig/diff_nh_sx_all_plus_M17_RCW38.eps}
{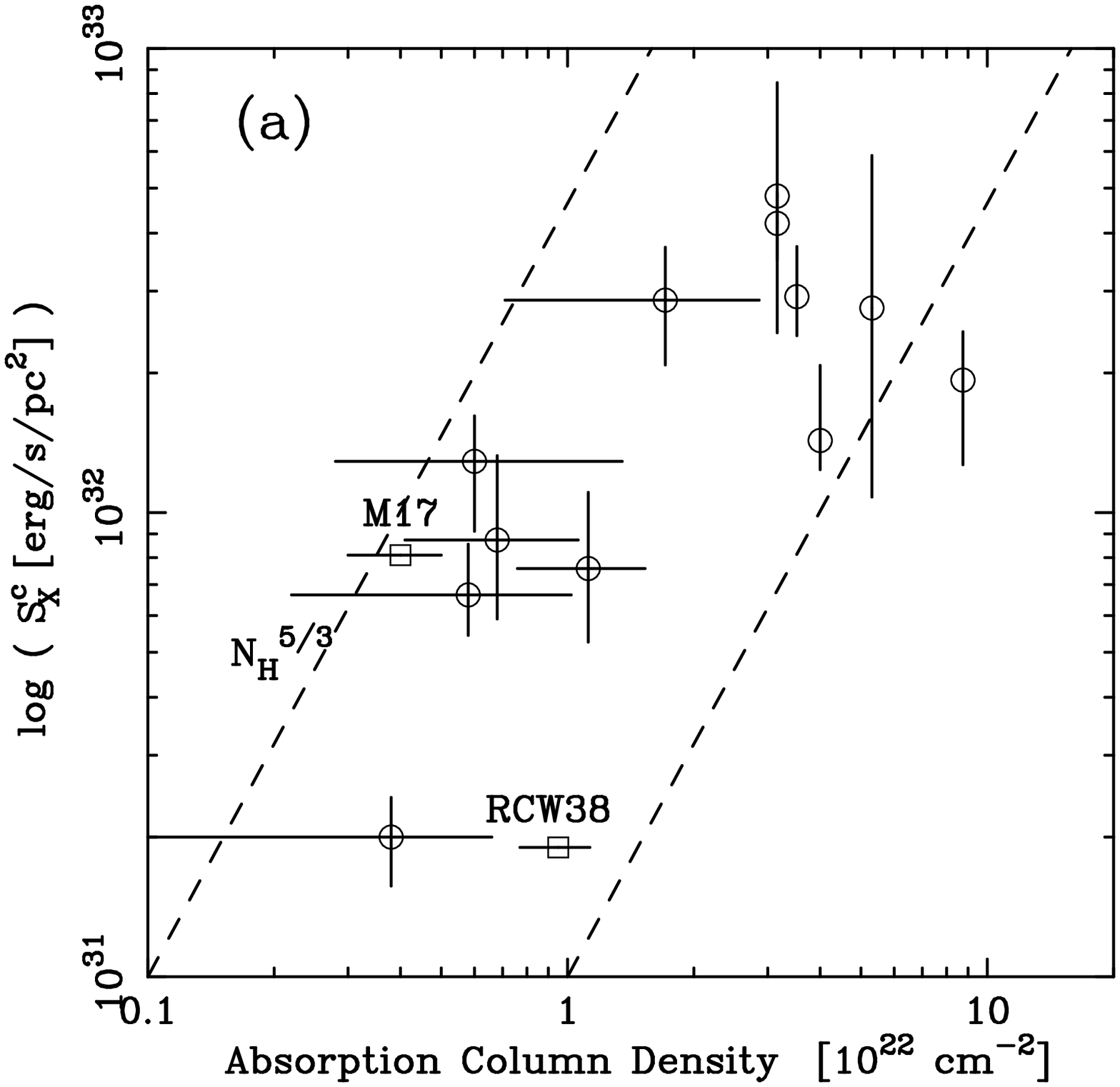}
\includegraphics[width=7cm,clip,angle=-90]
%%{fig/diff_nh_kt_plus_M17.eps}
{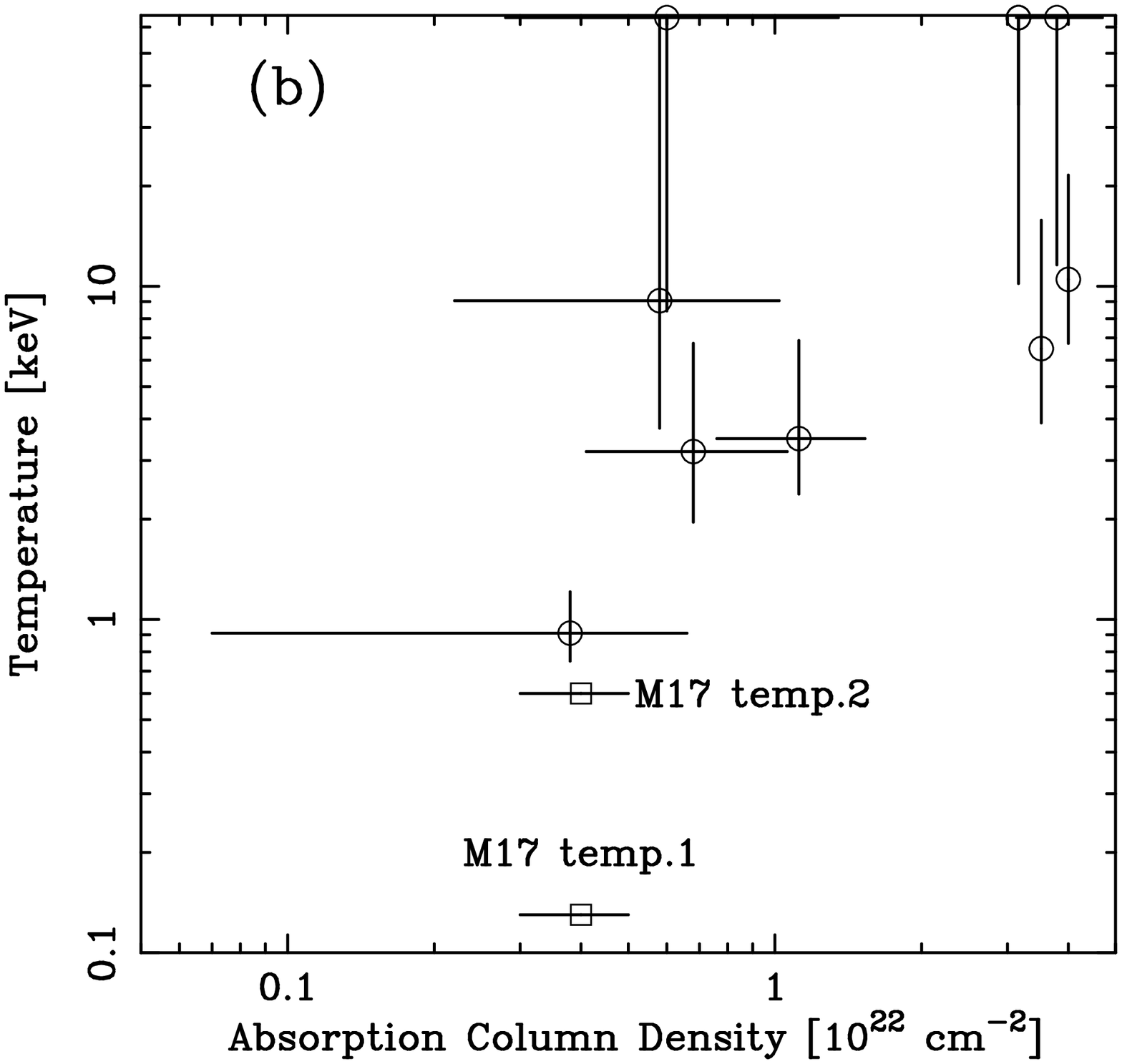}
}
%%\plottwo{f14a.eps}{f14b.eps}
\bigskip
\caption{(a) The relation between 
         the absorption column density and the absorption-corrected 
         0.5--8 keV surface brightness, for the 12 soft and 
         hard regions in NGC 6334.
         The dashed lines show 
         the prediction of equation (\ref{eqn:discuss:sx2}).
         (b) The absorption versus temperature diagram.
	 Squares show the reported results of M17 from 
	 \protect \cite{Townsley2003} 
	 and RCW 38 from \protect \cite{Wolk2002}. 
         }        
\label{fig:discuss:diff:summary}
\end{figure}

If the stellar wind and its confinement are
the origin of the observed diffuse X-ray emission,
we expect the emission properties to depend on the
confining pressure $p_{\rm s}$. Assuming in 
equation (\ref{eqn:discuss-wind-bubble-rb}) that 
$\dot{M}$, $v_{\rm w}$, and $t$ are unchanged,
we expect $R_{\rm b}\propto p_{\rm s}^{-\frac{1}{3}}$,
and hence $n_{\rm e}\propto \dot{M}t/R_{\rm b}^{3}\propto p_{\rm s}$,
where $n_{\rm e}$ is the density of the X-ray emitting plasma.
Since the X-ray volume emissivity scale as 
${n_{\rm e}}^2\propto{p_{\rm s}}^2$, 
we expect the absorption-corrected surface brightness to scale as 
\begin{equation}
\displaystyle
\begin{array}{cccccc}

S_{\rm X}^{\rm c} \propto {n_{\rm e}}^2 R_{\rm b} \propto {p_{\rm s}}^{2} \cdot {p_{\rm s}}^{-\frac{1}{3}} \propto {p_{\rm s}}^{\frac{5}{3}}.

\end{array}
\label{eqn:discuss:sx}
\end{equation}
Since the absorption column density $N_{\rm H}$ is an integrated
line-of-sight hydrogen density, we may very roughly assume 
$p_{\rm s} \propto N_{\rm H}$. This yields
\begin{equation}
\displaystyle
\begin{array}{cccccc}
S_{\rm X}^{\rm c}\propto{N_{\rm H}}^{\frac{5}{3}}.
\end{array}
\label{eqn:discuss:sx2}
\end{equation}
Indeed, as shown in figure \ref{fig:discuss:diff:summary} (a),
the observed surface brightness of the 12 regions shows a strong
positive correlation to the value of $N_{\rm H}$, and the dependence
is consistent with equation (\ref{eqn:discuss:sx2}).
This agreement holds even if excluding the hard regions of NGC 6334 
which is considered to be dominated by the non-thermal emission.
Although the lack of data points in the large absorption and low 
surface brightness region could be a selection effect, 
that in the small absorption and large surface 
brightness region is free from such artifacts. 

We similarly compared the absorption with the temperature 
in figure \ref{fig:discuss:diff:summary} (b).
This is equal to compare their continua hardness. 
Again we observe a positive correlation between the 
two quantities. Presumably, in dense environments, the stellar-wind 
shock becomes stronger. Consequently, the non-thermal emission may be 
dominant, or the shock temperature increases.

Thus, our {\it Chandra} result on NGC 6334 provide
a support to a view that the strong stellar winds from
young OB stars, confined by dense surrounding gas, give rise
to the diffuse hard X-ray emission. 

\subsection{Possible Contribution to the Galactic Ridge X-ray Emission}
\label{discuss:grxe}

Finally, we can roughly estimate the contribution of the diffuse emission
in galactic MSFRs. We here utilize the X-ray luminosity-to-mass ratio,
in order to estimate the contribution, following to \cite{Sekimoto2000}.
The total mass of the giant molecular cloud, birth places of massive
stars, in our Galaxies 
are estimated as $1\sim3\times10^{9}$ $\msolar$ from CO observations
\citep{Bronfman1988,Combes1991}.
The masses of NGC 6334 is estimated as $1.6\times10^5\msolar$
\citep{Dickel1977}. 
The mass to X-ray luminosity ratio of the diffuse emission becomes 
$1\times10^{28}$ \ergs $\msolar^{-1}$. If we take the mass of GMCs 
in our Galaxy as $2\times10^{9}$ $\msolar$, we obtain total X-ray 
luminosity of $2\times10^{37}$ \ergs.
This is $\sim10$\% of a hard 
tail of Galactic ridge X-ray emission 
($\sim2\times10^{38}$ \ergs in 2--10 keV, \citealt{Valinia1998}). 
Hence, a certain part of the Galactic ridge emission can be explained 
by diffuse emission in MSFRs.
                                                                             
\section{Conclusion}
\label{conclusion}

In the present paper, we have investigated the 
newly-suggested phenomenon of diffuse X-ray emission 
associated with the massive star formation NGC 6334
using the {\it Chandra} data. 
We have arrived at the following conclusions.

(1) After removing point sources, 
    the extended X-ray emission is detected with a high significance, 
    exhibiting a 0.5--8 keV luminosity of 2$\times10^{33}$ \ergs.
    It distributes over $\sim5\times9$ pc$^2$ and becomes bright in the
    vicinity of massive star-forming cores known in the optical, 
    infrared or radio wavelength ranges. The luminosity function within
    the extended emission suggests that most of the emission is diffuse 
    in nature.

(2) The diffuse emission outside the dense molecular cloud cores of NGC 6334
    show thermal spectra, with the temperature in the range of 1 to 10 keV.
    In the molecular cloud cores of NGC 6334, the emission exhibits
    very hard continua (photon indices of $\sim1$) with 
    significant absorption, 
    which prefer non-thermal interpretation to thermal scenario.
 
(3) The observed luminosity, temperature, possibly non-thermal emission, 
    and the angular extent of the emission are discussed in terms of shocks 
    which may be produced when fast stellar winds from 
    embedded young massive stars are confined by the thick materials 
    surrounding them.

The authors acknowledge technical advices from Dr. Tai Oshima,
Dr. Kensuke Imahishi and Dr. Masahiro Tsujimoto on the ACIS analysis.
They also thank Dr. Yasunobu Uchiyama and Prof. Takao Nakagawa for useful 
discussions and comments. YE is financially supported by the 
Japan Society for the Promotion of Science. 

\bibliographystyle{astron}
\bibliography{ms}

%%%%%%%%%%%%%%%%%%%%%%%%%%%%%%%%%%%%%%%%%%%%%%%%%%%%%%%%%%%%%%%%%%%%%%%%%%%%
%% figures
%%%%%%%%%%%%%%%%%%%%%%%%%%%%%%%%%%%%%%%%%%%%%%%%%%%%%%%%%%%%%%%%%%%%%%%%%%%%

%% fig. 1

%% fig. 2

%% fig. 3

%% fig. 4

%% fig. 5

%% fig. 6

%% fig. 7

%% fig. 8

%% fig. 9

%% fig. 10

%% fig. 11

%% fig. 12

%% fig. 13

%% fig. 14

\clearpage

%%%%%%%%%%%%%%%%%%%%%%%%%%%%%%%%%%%%%%%%%%%%%%%%%%%%%%%%%%%%%%%%%%%%%%%%%%%%
%% tables
%%%%%%%%%%%%%%%%%%%%%%%%%%%%%%%%%%%%%%%%%%%%%%%%%%%%%%%%%%%%%%%%%%%%%%%%%%%%

{\def\baselinestretch{1.2}
\begin{table}[htbp]
\caption{Parameters derived by the power-law model fitting to
         the EER spectrum.}
\label{tbl:ngc6334:chandra:diff:spec:param1}
\begin{center}
\begin{tabular}{cccccccccccccc} \hline
model  & param       &     \\\hline\hline
                                                                                
abs.$^a$   & $N_{\rm H}$ & 0.42$_{-0.18}^{+0.17}$ \\
P.L.$^b$   & $\Gamma$    & 0.85$_{-0.19}^{+0.17}$ \\
       & norm        & 4.4$_{-1.0}^{+1.1}\times10^{-4}$ \\
       & $F_{\rm X}$ & 5.6$\times10^{-12}$ \\
       & $L_{\rm X}$ & 2.2$\times10^{33}$ \\\hline
$\chi^2/\nu$ &       & 64.4 / 34  \\\hline
\end{tabular}
\end{center}
$^a$ Interstellar absorption, with $N_{\rm H}$ being
     the hydrogen column density in 10$^{22}$ cm$^{-2}$.\\
$^b$ Power-law model. $\Gamma$ is the photon index, norm is photon flux at 1 keV in photons cm$^{-2}$ s$^{-1}$, $F_{\rm X}$ and $L_{\rm X}$ are the X-ray flux
and absorption-corrected luminosity in the 0.5--8 keV band, respectively.
                                                                                
\end{table}
}

{\def\baselinestretch{1.2}
\begin{table}[htbp]
\caption{The same as table \ref{tbl:ngc6334:chandra:diff:spec:param1}, but 
         for the summed point-source spectrum fitted with the power-law plus 
         three narrow Gaussian models.}
\label{tbl:ngc6334:chandra:diff:spec:param2}
\begin{center}
\begin{tabular}{cccccccccccccc} \hline 
model  & param       &  \\\hline\hline

abs.$^a$   & $N_{\rm H}$    & 0.57$_{-0.02}^{+0.02}$ \\
P.L.$^b$   & $\Gamma$       & 1.25$_{-0.03}^{+0.03}$ \\
       & norm           & 1.28$_{-0.02}^{+0.02}\times10^{-3}$ \\

line1$^c$  & $E_{\rm C}$& 2.35$_{-0.01}^{+0.05}$ \\
       & $F_{\rm line}$ & 3.03$_{-0.65}^{+0.88}\times10^{-5}$ \\
       & $EW$           & 69 \\

line2$^c$  & $E_{\rm C}$& 3.13$_{-0.10}^{+0.02}$ \\
       & $F_{\rm line}$ & 1.85$_{-0.69}^{+0.45}\times10^{-5}$ \\
       & $EW$           & 60 \\

line3$^c$  & $E_{\rm C}$& 6.60$_{-0.04}^{+0.10}$ \\
       & $F_{\rm line}$ & 1.64$_{-0.74}^{+0.68}\times10^{-5}$ \\
       & $EW$           & 135 \\\hline

       & $F_{\rm X}$    & 9.6$\times10^{-12}$ \\\hline

$\chi^2/\nu$ &          & 256.4 / 170 \\\hline
\end{tabular}
\end{center}

$^a$ Interstellar absorption as defined in table \ref{tbl:ngc6334:chandra:diff:spec:param1}.\\
$^b$ Power-law model as defined in table \ref{tbl:ngc6334:chandra:diff:spec:param1}.\\
$^c$ A narrow Gaussian model. $E_{\rm C}$ is a line center energy in keV, 
$F_{\rm line}$ is line intensity in photons cm$^{-2}$ s$^{-1}$, and $EW$ 
is equivalent width in eV. \\

\end{table}
}

{\def\baselinestretch{1.2}
\begin{table}[htbp]
\caption{Results of the single temperature model fits to the soft-region spectra.}
\label{tbl:ngc6334:chandra:diff:spec:pos:param1}
\begin{center}
\begin{tabular}{cccccccccccccc} \hline
                                                                                                       
model        & param         & AXJ                                & C2                                & C3                               & C4E                                & C5N                                \\\hline\hline  
abs.$^a$     & $N_{\rm H}$   & 1.1$\pm0.4$                        & 0.38$_{-0.31}^{+0.28}$            & 0.60$_{-0.32}^{+0.75}$           & 0.68$_{-0.27}^{+0.38}$       & 0.58$_{-0.36}^{+0.44}$             \\		     
apec$^b$     & $kT^c$        & 3.5$_{-1.1}^{+3.4}$                & 0.91$_{-0.16}^{+0.30}$            & 64 ($>8.4$)                      & 3.2$_{-1.2}^{+3.6}$          & 9.1 ($>$3.8)                       \\		     
             & norm$^d$      & 3.4$_{-1.0}^{+1.6}$                & 0.41$\pm0.09$                     & 0.94$_{-0.28}^{+0.24}$           & 2.6$_{-0.8}^{+1.4}$          & 3.3$_{-0.6}^{+0.9}$                    \\		     
             & $F_{\rm X}^e$ & 2.2                                & 0.16                              & 0.98                             & 1.8                          & 3.8                                    \\		     
             & $L_{\rm X}^f$ & 1.4                                & 0.13                              & 0.42                             & 1.0                          & 1.8                                    \\\hline	     
$\chi^2/\nu$ &             & 55.5 / 45                          & 35.1 / 33                         & 20.7 / 20                        & 18.2 / 26                          & 31.4 / 35                          \\\hline        

\end{tabular}
\end{center}
$^a$ Interstellar absorption as defined in table \ref{tbl:ngc6334:chandra:diff:spec:param1}.\\
$^b$ The plasma emission model using the APEC (Astrophysical Plasma Emission Code), with the abundance fixed at 0.3 solar.\\
$^c$ $kT$ is a plasma temperature in keV.\\
$^d$ Norm is 10$^{-18}$/4$\pi D^2$ $EM$ or
     $EM$/(3.46$\times10^{51}$ cm$^{-3}$) where $D$ is a distance to NGC 6334
     and $EM$ is an emission measure in cm$^{-3}$.\\
$^e$ $F_{\rm X}$ is 0.5--8 keV X-ray flux in $10^{-13}$ \flux.\\
$^f$ $L_{\rm X}$ is 0.5--8 keV absorption-corrected luminosity in $10^{32}$ \ergs.\\
\end{table}
}

{\def\baselinestretch{1.2}
\begin{table}[htbp]
\caption{Results of the single temperature plasma model fits to the hard-region spectra$^a$.}
\label{tbl:ngc6334:chandra:diff:spec:pos:param2}
\begin{center}
\begin{tabular}{cccccccccccccc} \hline
                                                                                                       
model  &             & C1                                  & C1N                                  & C1S                              & C2W                                 \\\hline\hline
abs.   & $N_{\rm H}$ & 4.0 (fixed)                         & 8.8 (fixed)                          & 5.3  (fixed)                     & 3.5 (fixed)                         \\
apec   & $kT$        & 10$_{-3.8}^{+11}$                   & $>4.2$                               & 4.3$(>2.6)$                      & 6.5$_{-2.6}^{+9.3}$                 \\
       & $Z$         & 4.9$(>1.9)$                         & 0.3 (fixed)                          & 1.0$(<5.0)$                           & 0.3 (fixed)                         \\
       & norm        & 1.6$_{-0.2}^{+0.7}$                 & 1.0$_{-0.3}^{+0.3}$                  & 0.87$_{-0.53}^{+0.98}$           & 3.2$_{-0.6}^{+0.9}$                 \\
       & $F_{\rm X}$ & 2.7                                 & 0.58                                 & 0.53                             & 2.2                                 \\
       & $L_{\rm X}$ & 1.7                                 & 0.38                                 & 0.49                             & 1.6                                 \\\hline
$\chi^2/\nu$ &       & 21.4 / 28                           & 8.1 / 12                             & 14.5 / 14                        & 35.1 / 33                           \\\hline

       &             & C4                                   & CB                               & C4B                                 \\\hline\hline
abs.   & $N_{\rm H}$ & 3.2 (fixed)                          & 3.2  (fixed)                     & 3.8$_{-0.7}^{+0.9}$                 \\
apec   & $kT$        & $>10$                                & $>15$                            & $>12$                               \\
       & $Z$         & 0.3 (fixed)                          & 0.3  (fixed)                     & 0.3 (fixed)                         \\
       & norm        & 2.6$_{-0.8}^{+0.5}$                  & 2.7$_{-0.6}^{+0.5}$              & 1.1$\pm0.2$                         \\
       & $F_{\rm X}$ & 2.0                                  & 2.1                              & 8.5                                 \\
       & $L_{\rm X}$ & 1.2                                  & 1.2                              & 5.2                                 \\\hline
$\chi^2/\nu$ &       & 10.7 / 12                            & 32.3 / 26                        & 61.9 / 44                           \\\hline

\end{tabular}
\end{center}
$^a$ Notations and symbols are the same as table 
     \ref{tbl:ngc6334:chandra:diff:spec:pos:param1}, except $Z$ which represents the abundance in the solar unit.\\
\end{table}
}

{\def\baselinestretch{1.2}
\begin{table}[htbb]
\caption{Results of the power-law model fits to the C4, CB, and C4B spectra$^a$.}
\label{tbl:ngc6334:chandra:diff:spec:pos:param3}
\begin{center}
\begin{tabular}{cccccccccccccc} \hline

model      & param       & C4                                  & CB                                  & C4B                                  \\\hline\hline
abs.       & $N_{\rm H}$ & 3.2 (fixed)                         & 3.2 (fixed)                         & 1.7$_{-1.0}^{+1.2}$                  \\
P.L.       & $\Gamma$    & 1.0$_{-0.47}^{+0.44}$               & 0.56$_{-0.50}^{+0.49}$              & 0.39$_{-0.63}^{+0.66}$               \\
           & norm        & 2.8$_{-1.4}^{+2.1}$                 & 1.7$\pm0.3$                         & 4.2$_{-1.2}^{+1.3}$                  \\
           & $F_{\rm X}$ & 2.2                                 & 2.8                                 & 9.9                                  \\
           & $L_{\rm X}$ & 1.2                                 & 1.3                                 & 4.1                                  \\\hline
$\chi^2/\nu$ &           & 9.7/12                              & 32.3/26                             & 54.3/44                              \\\hline

\end{tabular}
\end{center}

$^a$ Notations and symbols are the same as table \ref{tbl:ngc6334:chandra:diff:spec:param1},
     except that norm is in 10$^{-5}$ photons cm$^{-2}$ s$^{-1}$, $F_{\rm X}$ is in $10^{-13}$ \flux, 
     and $L_{\rm X}$ is in $10^{32}$ \ergs.\\

\end{table}
}

{\def\baselinestretch{1.2}
\begin{table}[htbb]
\caption{Results of a leaky-absorber model fit to the C4B spectrum$^a$.}
\label{tbl:ngc6334:chandra:diff:spec:pos:param4}
\begin{center}
\begin{tabular}{cccccccccccccc} \hline

model      & param       &\\\hline\hline
abs.1      & $N_{\rm H}$ &4.0(fixed)                           \\
apec1      & norm        &11$_{-2}^{+1}$        \\
           & $F_{\rm X}$ &8.0                   \\
           & $L_{\rm X}$ &4.9                   \\\hline
abs.2      & $N_{\rm H}$ &$<0.92$                              \\
apec2      & $kT$        &$>30$                                \\
           & norm        &0.42$_{-0.24}^{+0.77}$     \\
           & $F_{\rm X}$ &0.5                  \\
           & $L_{\rm X}$ &1.9                  \\\hline

$\chi^2/\nu$ &           &53.7/43                              \\\hline

\end{tabular}
\end{center}
\begin{tabular}{l}
$^a$ Notations and symbols are the same as table \ref{tbl:ngc6334:chandra:diff:spec:pos:param1}.
     Abundance is fixed at 0.3 solar.
\end{tabular}
\end{table}
}

\end{document}